%% file: main.tex
\documentclass[sigconf]{acmart}
\usepackage{graphicx} % Required for inserting images
\usepackage{booktabs}
\usepackage{multirow}
\usepackage{makecell}
\usepackage{colortbl}
\usepackage{array}
\usepackage{siunitx}
\usepackage{amsmath}
\usepackage{enumitem}
\usepackage{subfigure}
\usepackage{pifont}
\setlist[enumerate]{align=left}
\newcommand{\cmark}{\ding{51}}
\newcommand{\xmark}{\ding{55}}
\AtBeginDocument{%
  }

\copyrightyear{2026}
\acmYear{2026}
\setcopyright{cc}
\setcctype{by}
\acmConference[CIKM '26]{Proceedings of the 35th ACM International Conference on Information and Knowledge Management}{November 07--11, 2026}{Rome, Italy}
\acmBooktitle{Proceedings of the 35th ACM International Conference on Information and Knowledge Management (CIKM '26), November 07--11, 2026, Rome, Italy}
\acmDOI{10.1145/3799682.3841085}
\acmISBN{979-8-4007-2539-5/2026/11}

\begin{document}
\title{Retrieve-then-Adapt: Test-Time Retrieval-Augmentation for Sequential Recommendation}

\author{Xing Tang}
\authornote{These authors contributed equally to this research.}
%\orcid{1234-5678-9012}
\affiliation{%
  \institution{School of Artificial Intelligence Shenzhen Technology University}
  \city{Shenzhen}
  \country{China}
}

\author{Ziqiang Cui}
\authornotemark[1]
\affiliation{
  \institution{City University of Hong Kong}
  \city{Hong Kong}
  \country{Hong Kong SAR}
}

\author{Jingyang Bin}
%\authornotemark[1]
\affiliation{%
  \institution{Shenzhen Technology University}
  \city{Shenzhen}
  \country{China}
}

\author{Xiaokun Zhang}
\affiliation{
  \institution{City University of Hong Kong}
  \city{Hong Kong}
  \country{Hong Kong SAR}
}
\author{Fuyuan Lyu}
\affiliation{
  \institution{McGill University}
  \city{Montreal}
  \country{Canada}
}

\author{Jingyan Jiang}
\affiliation{ 
\institution{Shenzhen Technology University} \city{Shenzhen} 
\country{China} 
}

\author{Dugang Liu}
%\authornote{Corresponding authors.}
\affiliation{
  \institution{Shenzhen University}
  \city{Shenzhen}
  \country{China}
}

\author{Chen Ma}
\affiliation{
  \institution{City University of Hong Kong}
  \city{Hong Kong}
  \country{Hong Kong SAR}
}

\author{Ruiming Tang}
\affiliation{
  \institution{Kuaishou Technology}
  \city{Shenzhen}
  \country{China}
}

\author{Xiuqiang He}
\authornote{Corresponding author}
\affiliation{
  \institution{Shenzhen Technology University}
  \city{Shenzhen}
  \country{China}
}

\renewcommand{\shortauthors}{Xing Tang, et al.}
\input{section/abstract}



  
\maketitle

\input{section/introduction}
\input{section/related}
\input{section/preliminary}
\input{section/method}
\input{section/experiment}

\input{section/conclusion}
\input{section/acknowledgments}

\clearpage
\input{section/genai_disclosure}
\bibliographystyle{ACM-Reference-Format}
\bibliography{reference}
\end{document}

%% file: section/abstract.tex
\begin{abstract}

The sequential recommendation (SR) task aims to predict the next item based on users' historical interaction sequences. Typically trained on historical data, SR models are frozen after deployment, limiting their ability to handle test-time patterns that are rare during training.
Existing approaches to address this limitation include test-time training, test-time augmentation, and retrieval-augmented fine-tuning. However, these methods either introduce significant computational overhead, rely on random augmentation strategies, or require a carefully designed two-stage training paradigm.
In this paper, we argue that the key to effective test-time augmentation lies in achieving both effective retrieval and efficient fusion. To this end, we propose Retrieve-then-Adapt (\textbf{ReAd}), a novel framework that augments a deployed SR model through retrieved user preference signals. Specifically, given a trained SR model, ReAd first retrieves collaboratively similar items for a test user from a constructed collaborative memory. A lightweight retrieval learning module then integrates these items into an informative augmentation embedding that captures both collaborative signals and prediction-refinement cues. Finally, the initial SR prediction is refined via a fusion mechanism that incorporates this embedding. Extensive experiments across five benchmark datasets demonstrate that ReAd consistently outperforms existing SR methods. The source code is publicly available at \url{https://github.com/xingt-tang/ReAd}.

\end{abstract}

%% file: section/introduction.tex
\section{Introduction}
\label{sec:intro}

In recent years, sequential recommendation (SR) has attracted considerable attention for its ability to learn user preferences from sequential behavioral data~\cite{hstu,onerec}. Early SR research primarily focused on designing increasingly sophisticated architectures to capture implicit user patterns from historical interactions~\cite{GRU4Rec,bert4rec,sasrec,caser,CGN,SR-GNN}. However, due to the limited number of interactions between users and items, training SR models often suffers from data sparsity, making it challenging to learn high-quality user representations~\cite{meta}. To mitigate this issue, some recent studies have turned to self-supervised learning (SSL), leveraging data augmentation and contrastive loss to align different augmented views of the data~\cite{Cadirec}. With the success of contrastive learning, SR models can now incorporate additional knowledge during training, thereby enhancing the quality of user representations.

\begin{figure}[!t]
    \centering
    \includegraphics[width=0.98\linewidth]{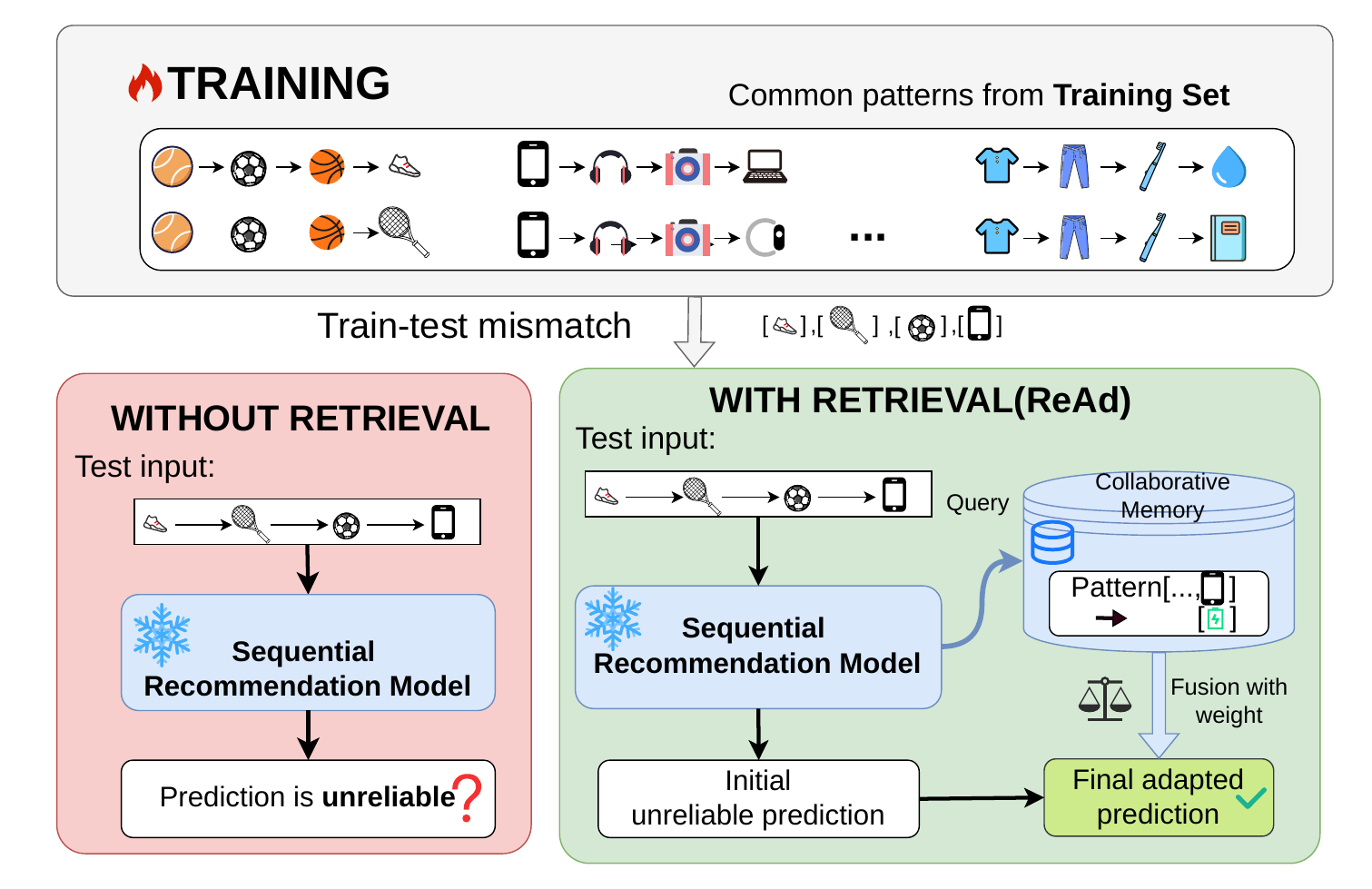}
    \caption{Illustration of the training-test mismatch in sequential recommendation. The training set contains common patterns (e.g., sports or electronics sequences). However, the test input may contain items or combinations that are rare in the training data. Without retrieval (left), the frozen model produces unreliable predictions. With ReAd (right), the test sequence queries a collaborative memory, retrieves relevant patterns, and fuses them with the initial prediction to produce an adapted output.}
    \Description{A two-part diagram contrasts frozen sequential recommendation without retrieval against ReAd. ReAd uses a rare mixed test sequence to query a collaborative memory, retrieves relevant historical patterns, and fuses them with the initial prediction.}
    \label{fig:intro}
\end{figure}

Despite efforts to train the SR model, improving its performance at test time or the inference stage remains an open challenge. As illustrated in Figure~\ref{fig:intro}, deployed SR models rely on frozen parameters inherited from training. This static representation poses challenges when test-time patterns differ from those seen during training.
First, training data is often dominated by popular items and frequent interaction patterns, while test-time sequences may contain long-tail items or rare combinations~\cite{TTT,DTT}. This mismatch between training and test distributions is a fundamental challenge for frozen models. For instance, consider the example in Figure~\ref{fig:intro}: during training, the model frequently encounters patterns such as \texttt{[tennis racket, soccer ball, basketball] -> sneakers}. However, at test time, the input sequence \texttt{[sneakers, tennis racket, soccer ball, smartphone]} contains a smartphone, which rarely appears in such sports-related combinations in the training set.
Second, trained models encode collaborative signals into their parameters, which may be insufficient for rare patterns. In particular, interactions involving long-tail items often lack adequate representation and are easily dominated by popular items. When encountering a test-time sequence, SR models rely heavily on fixed parameters that encode historical collaborative signals. However, these signals may not generalize to long-tail items or rare combinations, leading to a mismatch between training and test distributions. Together, these challenges highlight a fundamental issue in SR, how to help a frozen model generalize to test-time patterns that are rare or absent in training.

Recent studies have explored test-time learning paradigms to improve sequential recommendation during inference. Test-time training has been introduced into recommendation systems~\cite{TTT,DTT,ttt4rec}, leveraging an auxiliary task to facilitate real-time model updating. However, as SR architectures become increasingly complex~\cite{hstu,fuxi-alpha,longer}, the computational overhead of such online updates can no longer be overlooked during deployment. Alternatively, lightweight test-time augmentation methods have been explored, which enhance sequential recommendation by augmenting input sequences at inference time and aggregating predictions without introducing additional tasks or structures~\cite{tta,empowering}. Yet, because it relies on random augmentation operators and simple averaging, this method often lacks robustness and may fail to generalize effectively. Another line of work addresses training-test distribution mismatch through retrieval-augmented fine-tuning, which accelerates adaptation by combining recommendation and retrieval learning during pre-training~\cite{Raserec}. This approach, however, requires a carefully structured two-stage training pipeline that jointly incorporates both objectives. Overall, there remains a clear need for a low-cost, efficient, and model-agnostic approach that effectively enhances sequential recommendation performance at test time.

In this paper, we propose a novel framework, \textbf{Re}trieve-then-\textbf{Ad}apt (ReAd), to address the challenges outlined above. Unlike prior test-time adaptation methods that update parameters or rely on random augmentation, ReAd performs test-time retrieval augmentation while keeping the SR model completely frozen. Designing such a retrieval-augmented framework for sequential recommendation presents two main challenges. First, unlike language models~\cite{rag} or vision-language models~\cite{RA-TTA}, SR lacks access to external knowledge sources for retrieval-based enhancement. Defining a suitable knowledge base is thus crucial. We must identify what knowledge is essential and establish a feasible retrieval mechanism. As noted earlier, collaborative signals encoded within model parameters are often weak, especially for long-tail items. To this end, we construct a collaborative memory from the training data that maps sequence representations to their next items. This design enables the retrieval of relevant items based on historically similar behavior sequences, thereby augmenting predictions explicitly at test time—without relying solely on implicit parametric knowledge.
The second challenge lies in effectively integrating the retrieved items. Similar to retrieval-augmented generation~\cite{rag}, which retrieves multiple documents per query, our method also retrieves several items per query. To enable efficient augmentation for each test sample, we aim to refine predictions directly from these retrieved items without incurring high computational costs. To this end, we introduce a lightweight retrieval learning module that fuses the retrieved items into a unified representation via cross-attention, coupled with an entropy-based fusion mechanism that dynamically balances the initial prediction and the retrieved signal.

We summarize our major contributions as follows:
\begin{itemize}[topsep=0pt,noitemsep,nolistsep,leftmargin=*]
\item We systematically study test-time retrieval augmentation for sequential recommendation, characterizing when and why it helps and revealing that its effectiveness correlates with interaction density and item popularity.

\item We propose ReAd, a lightweight framework that retrieves relevant patterns from a collaborative memory at test time and fuses them with frozen-model predictions, without parameter updates or retraining.

\item Extensive experiments are conducted on five public datasets. The experimental results demonstrate the effectiveness and efficiency of the proposed method.
\end{itemize}

The remainder of this paper is structured as follows. Section~\ref{sec:rw} reviews related work in sequential recommendation, test-time learning paradigms, and retrieval-augmented recommendation. Section~\ref{sec:pre} introduces the preliminary definitions and task formulation for sequential recommendation. Our proposed ReAd framework is presented in detail in Section~\ref{sec:method}, followed by experimental results and analysis in Section~\ref{sec:exp}. Finally, Section~\ref{sec:con} concludes the paper and outlines potential directions for future work.

%% file: section/related.tex
\section{Related Work}
\label{sec:rw}

In this section, we provide a brief review of related work. Our study builds upon three key research lines: sequential recommendation, test-time strategies for sequential recommendation, and retrieval-augmented recommendation.

\begin{table*}[!htbp]
\centering
\small
\renewcommand{\arraystretch}{0.90}
\begin{tabular}{lcccccc}
\toprule
\textbf{Method} & 
\multicolumn{1}{c}{\textbf{Test-time}} & 
\multicolumn{1}{c}{\textbf{Retrieval}} & 
\multicolumn{1}{c}{\textbf{Test-time}} & 
\multicolumn{1}{c}{\textbf{Requires}} & 
\textbf{Plug-and-play} & 
\multicolumn{1}{c}{\textbf{Computational}} \\
& 
\multicolumn{1}{c}{\textbf{operation}} & 
\multicolumn{1}{c}{\textbf{enhanced}} & 
\multicolumn{1}{c}{\textbf{param update}} & 
\multicolumn{1}{c}{\textbf{retraining}} & 
& 
\multicolumn{1}{c}{\textbf{overhead}} \\
\midrule
TTT4Rec & \cmark & \xmark & \cmark & \xmark & \xmark & High \\
TTA & \cmark & \xmark & \xmark & \xmark & \cmark & Low \\
RaSeRec & \xmark & \cmark & \xmark & \cmark & \xmark & High \\
\textbf{ReAd (Ours)} & \cmark & \cmark & \xmark & \xmark & \cmark & Low \\
\bottomrule
\end{tabular}
\caption{Learning paradigms for sequential recommendation: test-time training, test-time augmentation, training-time retrieval, and ReAd.}
\label{tab:comparison}
\end{table*}

\subsection{Sequential Recommendation}
The learning paradigm of sequential recommendation (SR) models has progressed from traditional models~\cite{fusing,markov} to deep learning approaches~\cite{GRU4Rec,bert4rec,sasrec,caser,mlp}. For instance, GRU4Rec~\cite{GRU4Rec} and Caser~\cite{caser} employed Gated Recurrent Units (GRU) and convolutional networks, respectively, to capture the dynamics of user preferences. Following the success of attention mechanisms in language modeling~\cite{attention}, transformer-based SR models have gained prominence~\cite{sasrec,bert4rec}. More recently, inspired by large foundation models in language processing, several studies have explored scaling up the parameter sizes of SR models~\cite{hstu,fuxi-alpha}. 
Another important research direction addresses data sparsity in sequential recommendation by integrating self-supervised learning (SSL) techniques~\cite{s3,cl4srec,duorec,coserec}. A key aspect of these approaches is data augmentation, which has evolved from heuristic operations—such as cropping, masking, and dropout—to model-enhanced strategies, including diffusion models~\cite{Cadirec, Diff4rec} and large language model-based augmentation~\cite{SRA-CL}. Additional efforts focus on improving semantic alignment across augmented views through techniques such as contrastive learning and invariant representation learning~\cite{equivariant,ICL,mcl}. 

Despite these advances, existing methods primarily operate during training and may struggle with test-time patterns that are rare in the training data, limiting their effectiveness in real-world recommendation scenarios.

\subsection{Test-Time Learning Paradigms for SR}
Building on the preceding discussion, recent research has increasingly focused on test-time strategies for sequential recommendation, particularly test-time training (TTT)~\cite{ttt++} and test-time augmentation (TTA)~\cite{better}.
For example, TTT4Rec~\cite{ttt4rec} adapts model parameters during inference by leveraging additional real-time data. Following this direction, T$^2$ARec~\cite{TTT} and PCRec~\cite{breaking} further explore ways to update trained models at test time. T$^2$ARec introduces a state-space model with two alignment modules to capture shifts in user interest distributions, while PCRec incorporates real-time hidden-state inference and performs one-step optimization during deployment. Despite their effectiveness, these approaches require specialized architectures and incur non-negligible computational overhead during inference. In parallel, TTA-based operators such as TNoise and TMask~\cite{tta} investigate data augmentation applied directly at test time. While these enhancements yield performance gains over baseline models, they remain susceptible to failure due to the inherent randomness of augmentation and potential misalignment with model architecture.
\subsection{Retrieval-Augmented Recommendation}

Retrieval-Augmented Generation (RAG) has gained prominence in large language models (LLMs) for its ability to incorporate external knowledge from structured databases ~\cite{rag,rag-survey}. Inspired by this paradigm, several works have introduced retrieval-augmented strategies into recommendation systems to enhance performance~\cite{ReDA,Raserec,SRA-CL,CoRAL,RAL-CDNet,RALLRec}. These approaches retrieve external knowledge from diverse sources, such as LLM-generated content ~\cite{SRA-CL,CoRAL}, cross-domain information ~\cite{RAL-CDNet}, or historical user behavior ~\cite{Raserec}. Among these, RaSeRec~\cite{Raserec} is particularly relevant to our work. It proposes a two-stage framework involving pre-training and fine-tuning to mitigate training-test mismatch. However, like other similar methods, it operates primarily during the training phase and cannot adapt dynamically during inference without retraining, thus limiting its applicability in real-time recommendation settings.

To better understand the differences among these paradigms, Table~\ref{tab:comparison} summarizes the key characteristics of test-time training (TTT4Rec), test-time augmentation (TTA), training-time retrieval (RaSeRec), and our proposed ReAd.

%% file: section/preliminary.tex
\section{Preliminaries}
\label{sec:pre}
In this section, we first introduce the standard sequential recommendation model, followed by the formal problem formulation for test-time retrieval augmentation.
\subsection{Sequential Recommendation Model}

We begin with the notation used in sequential recommendation. The goal of sequential recommendation is to predict the next item a user is likely to interact with based on their historical behavior sequence. Let $\mathcal{U} = \{u_1, u_2, \dots, u_{|\mathcal{U}|}\}$ denote the set of users and $\mathcal{V} = \{v_1, v_2, \dots, v_{|\mathcal{V}|}\}$ denote the set of items, where $|\mathcal{U}|$ and $|\mathcal{V}|$ are the number of users and items, respectively. For each user $u \in \mathcal{U}$, the historical interaction sequence in chronological order is represented as $\mathbf{s}^u = \{v^u_1, v^u_2, \dots, v^u_{|\mathbf{s}^u|}\}$, where $v^u_t \in \mathcal{V}$ is the item that user $u$ interacted with at time step $t$.
Given $\mathbf{s}^u$, a sequential recommendation model $M$ is trained to maximize the likelihood of the next item:

\begin{equation}
\arg\max_{v^* \in \mathcal{V}} p\bigl(v_{|\mathbf{s}|+1} = v^* \mid \mathbf{s}^u\bigr),
\end{equation}
where $p(\cdot \mid \mathbf{s}^u)$ denotes the output probability distribution of the model, representing the likelihood of candidate items given the user's historical sequence.

\subsection{Problem Formulation}

At test time, our objective is to augment the trained sequential model $M$ to improve next-item prediction for a given user $u_t$ and her input sequence $\mathbf{s}^{u_t}$, while leveraging item embeddings $\{\mathbf{e}_j \mid j \in \mathcal{V}\}$. To enable retrieval-based augmentation, a key subproblem is to construct an indexed collaborative memory $\mathcal{D}$ that supports efficient lookup of relevant items. Formally, the augmentation step can be expressed as:

\begin{equation}
f_{\text{aug}}: (\mathcal{D},M; \mathbf{s}^{u_t}) \longrightarrow \mathbf{e}_{\text{aug}},
\end{equation}
where $f_{\text{aug}}$ is a retrieval-augmentation function that maps the memory $\mathcal{D}$ and the input sequence $\mathbf{s}^{u_t}$ to an augmented embedding $\mathbf{e}_{\text{aug}}$.
Based on this augmented representation, the final prediction refinement follows:

\begin{equation}
v^* = g_{\text{refine}}\Bigl(p_{\text{init}},p_{\text{aug}}\Bigr),
\label{eq:refine}
\end{equation}
where $g_{\text{refine}}$ denotes the refinement operator applied at test time, and $v^*$ is the final predicted item based on the initial prediction $p_{\text{init}}$ and the augmented prediction $p_{\text{aug}}$.

Note that during test-time inference, the original training data is not accessible, and the forward computation through $M$ is assumed to be computationally efficient. The core contributions of this paper focus on designing effective instantiations of the two functions $f_{\text{aug}}$ and $g_{\text{refine}}$.

%% file: section/method.tex
\section{Methodology}
\label{sec:method}

\begin{figure*}[!htbp]
    \centering
    \includegraphics[width=0.96\linewidth]{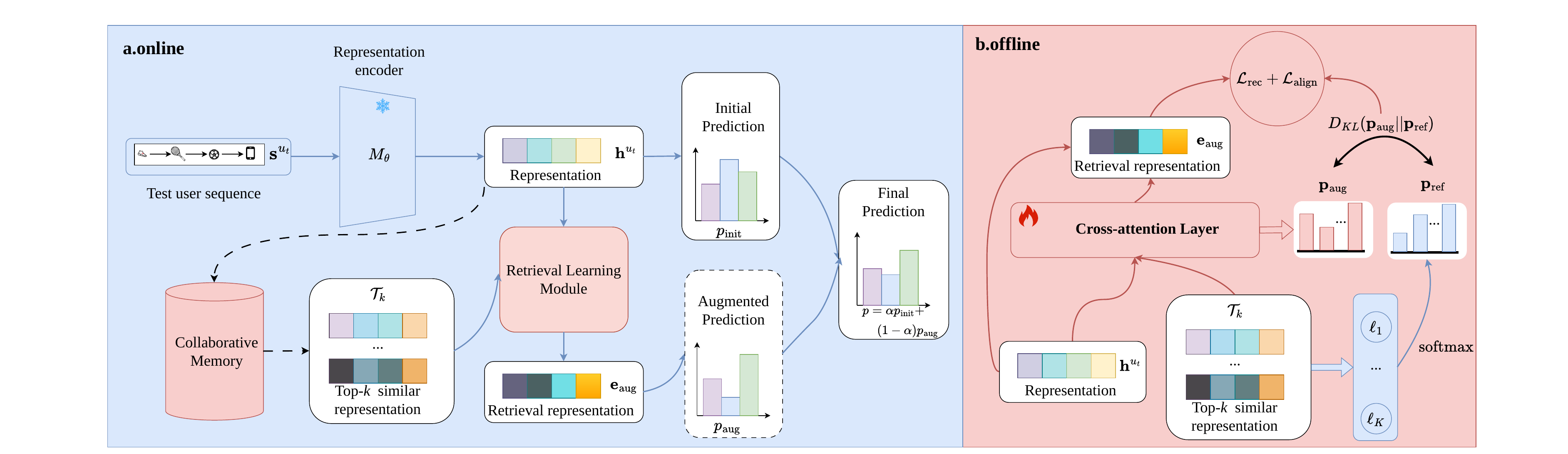}
    \caption{Overview of ReAd. (a) Online inference retrieves collaborative memories and fuses them with the initial prediction. (b) Offline retrieval learning trains cross-attention with recommendation and KL-based alignment losses. Blue and red boxes denote online and offline components, respectively.}
    \Description{A two-panel flowchart of ReAd. The online panel retrieves similar representations from collaborative memory and fuses them with the initial prediction. The offline panel trains a cross-attention retrieval module with recommendation and alignment losses.}
    \label{fig:framework}
\end{figure*}

In this section, we present the proposed \textbf{ReAd} (Retrieve-then-Adapt) framework in detail. ReAd is designed as a test-time retrieval augmentation method. It does not modify the pre-trained SR model's parameters. Instead, it constructs an offline collaborative memory of training patterns and retrieves from it at test time to augment the model's prediction. We begin with a high-level overview of its workflow, followed by a step-by-step explanation of the retrieval-augmentation and refinement mechanisms in subsequent subsections.

\subsection{The Overview of ReAd}
Figure~\ref{fig:framework} presents the overall framework of the proposed ReAd method. The framework operates in two main steps: offline preparation (completed once before 
deployment) and online inference (forward pass only, no gradient updates).

In the offline stage (Section~\ref{sec:cra}), we construct a collaborative memory $\mathcal{D}$ from the training data. To effectively integrate the retrieved candidate items, we design a retrieval learning module that processes candidate item embeddings. Notably, the parameters of the trained sequential model remain fixed during this stage, ensuring that the retrieval process remains lightweight and does not introduce additional training overhead.

During the online inference stage (Section~\ref{sec:tta}), for a given test user sequence, the retrieval module selects the top-$k$ most relevant items from $\mathcal{D}$ based on sequence similarity and fuses their embeddings into a retrieved representation. The parameters of the SR model remain frozen throughout; no gradient updates are performed at test time. The retrieved representation is then used to refine the model's initial prediction, ultimately yielding the final augmented prediction through a learnable fusion mechanism.

\subsection{Collaborative Memory Construction}
\label{sec:cra}

During the training stage, the sequential recommendation model $M$ is optimized using historical user behavior data to learn and encode collaborative signals. To provide necessary context for the subsequent adaptation process, we first outline the standard training procedure.

\subsubsection{Model Training}

We construct a trainable item embedding matrix $M_{\mathbf{E}}\in \mathbb{R}^{|\mathcal{V}| \times d}$ for the entire item set $\mathcal{V}$, 
mapping each item to a $d$-dimensional dense vector. During training, each user sequence $\mathbf{s}^u$ is split into subsequence–label pairs $(\{v_1^u, \cdots,v_j^u\},v_{j+1}^u)$, where the latter denotes the item to be predicted.
Given a subsequence $\{v_1^u, \cdots,v_j^u\}$, the sequential model $M$ first computes its representation $\mathbf{h}_j^u\in \mathbb{R}^{1\times d}$ via the representation encoder, denoted $M_\theta$. The similarities between $\mathbf{h}_j^u$ and all item embeddings are then obtained as:
\begin{equation}
\label{eq:similarity}
\mathbf{r} = \mathbf{h}_j^u M_\mathbf{E}^\top,
\end{equation}
where $\mathbf{r}\in \mathbb{R}^{|\mathcal{V}|}$, and the value of each entry $r_i$ indicates the unnormalized affinity between the sequence representation and item $v_i$, and the ranking of $r_i$ determines the predicted order of candidate items. While inference relies on the top-$k$ ranking derived from $\mathbf{r}$, training optimizes the negative log-likelihood over the full item set via softmax:

\begin{equation}
\label{eq:rec}
\mathcal{L}_{\text{train}}=-\sum_{u\in\mathcal{U}}\sum_{j=1}^{|\mathbf{s}^u|}\log(e^{\mathbf{h}_j^u\mathbf{e}_{v_{j+1}^u}^\top}/\Sigma_{v_i\in\mathcal{V}}e^{\mathbf{h}_j^u\mathbf{e}_{v_i}^\top}).
\end{equation}
Here, $\mathbf{e}_{v_{j+1}^u}$ denotes the embedding of the target item $v_{j+1}^u$, and $\mathbf{e}_{v_i}$ represents the embedding of any item in $\mathcal{V}$. The model parameters are updated by minimizing $\mathcal{L}_{\text{train}}$, thereby encoding collaborative signals in a parametric manner.

\subsubsection{Collaborative Memory}

Based on the trained representation encoder $M_{\theta}$, we construct a collaborative memory $\mathcal{D}$ that explicitly stores collaborative signals. Specifically, for each training sequence $\mathbf{s}^u$, we compute its representation $\mathbf{h}^u$ using $M_{\theta}$. The embedding of the corresponding next item $\mathbf{e}_{v_u}$ is associated with $\mathbf{h}^u$ to form a pair $<\mathbf{h}^u,\mathbf{e}_{v_u}>$, which encapsulates a sequential pattern in user behavior. Here, $\mathbf{h}^u$ serves as the index in $\mathcal{D}$, and the collection of such pairs comprises the entries of the memory. Sequences with similar representations $\mathbf{h}$ typically correspond to similar target items, reflecting that collaborative signals are explicitly encoded and accessible in this memory structure.

During inference, given a test user sequence $\mathbf{s}^{u_t}$, we first derive its representation $\mathbf{h}^{u_t}$ via $M_\theta$. This representation is then used to retrieve the top-$k$ most relevant item embeddings from $\mathcal{D}$ based on cosine similarity:

\begin{align}
\label{eq:topk}
&\mathcal{T}_K=\{\mathbf{e}_k|s_{u_t,k}\in \operatorname{Top}-k(\mathbf{h}^{u_t},\mathcal{D},K)\},\\
&\text{where}\quad s_{u_t,k}=\cos(\mathbf{h}^{u_t},\mathbf{h}^u)=\frac{\mathbf{h}^{u_t}\cdot\mathbf{h}^u}{\Vert \mathbf{h}^{u_t}\Vert\Vert\mathbf{h}^u\Vert}, \notag
\end{align}
where $\operatorname{Top}-k(.;K)$ is a function that selects the $K$ highest values from the memory regarding the representation as a query. For efficiency, the retrieval process can be accelerated using approximate nearest-neighbor search libraries such as FAISS~\cite{faiss}.

\subsubsection{Retrieval Learning}

To utilize the retrieved item embeddings from $\mathcal{T}_K$, a simple strategy is to average predictions based on Equation~\ref{eq:similarity}, which is commonly employed in test-time augmentation methods~\cite{tta}. While efficient, this strategy introduces increasing noise as $K$ grows. Another straightforward alternative is to manually weight the embeddings, which is both subjective and difficult to generalize.

We propose a lightweight, learning-based weighting scheme, illustrated in the offline part of Figure~\ref{fig:framework}, which requires minimal additional computation during retrieval. Specifically, we treat the sequence representation $\mathbf{h}^{u_t}$ as the query and the retrieved item set $\mathcal{T}_K$ as both the keys and the values. The retrieved embedding is then obtained via cross-attention:

\begin{align}
\label{eq:atten}
&\mathbf{e}_{\text{aug}} =\operatorname{softmax}(\frac{QK^\top}{\sqrt d})V, \notag &\\
&\text{where}\quad Q=\mathbf{h}^{u_t}\mathbf{W}_q,K=\mathbf{e}_K\mathbf{W}_k,V=\mathbf{e}_K\mathbf{W}_v.
\end{align}
Here, $\mathbf{e}_K\in \mathbb{R}^{K\times d}$ stacks the retrieved embeddings, and $\mathbf{W}_q\in \mathbb{R}^{d\times d'}$, $\mathbf{W}_k\in \mathbb{R}^{d\times d'}$ and $\mathbf{W}_v\in \mathbb{R}^{d\times d}$ are learnable projection matrices. The attention distribution over the retrieved items is given by:

\begin{equation}
\label{eq:paug}
\mathbf{p}_{\text{aug}}= \operatorname{softmax}(\frac{QK^\top}{\sqrt d}),
\end{equation}
which reflects the similarity between the query and each retrieved item under the learned transformations. 

The projection matrices are optimized through two complementary loss terms: a recommendation loss that ensures the augmented representation improves prediction accuracy, and an alignment loss that calibrates the attention weights based on each retrieved item's intrinsic predictive utility.

We define the combined representation as the average of the query representation 
and the augmented embedding: $\mathbf{h} = \frac{1}{2}(\mathbf{h}^{u_t} + 
\mathbf{e}_{\text{aug}})$, and train it to better predict the target item 
$\mathbf{e}_t$.
This simple equal weighting suffices, as the subsequent entropy-based fusion mechanism at test time will dynamically adjust the final contribution.
The recommendation loss then ensures the augmented representation to improve the prediction:
\begin{equation}
\mathcal{L}_{\text{rec}} = -\log \frac{\exp\left(\mathbf{h} \mathbf{e}_t^\top\right)}{\sum_{v_i \in \mathcal{V}} \exp\left(\mathbf{h} \mathbf{e}_{v_i}^\top\right)}.
\end{equation}

While $\mathbf{p}_{\text{aug}}$ in Equation~\ref {eq:paug} captures similarity between the query and retrieved items, this similarity alone may not reflect an item's true utility for refining the final prediction. For instance, a highly similar popular item could provide redundant information, whereas a moderately similar but discriminative long-tail item might be more beneficial for correcting the prediction. To address this gap, we introduce an alignment loss, which calibrates $\mathbf{p}_{\text{aug}}$ by aligning it with a reference distribution $\mathbf{p}_{\text{ref}}$ that explicitly encodes each retrieved item's predictive usefulness. 
We first measure how well each retrieved item $\mathbf{e}_k$ independently 
predicts the target item $\mathbf{e}_t$:
\begin{equation}
\ell(\mathbf{e}_k,\mathbf{e}_t)= -\log \frac{(\exp({\mathbf{e}_k\mathbf{e}^\top_{t})}}{\Sigma_{v_i\in \mathcal{V}}\exp({\mathbf{e}_k\mathbf{e}^\top_{v_i}})}.
\end{equation}
A lower $\ell(\cdot)$ indicates stronger predictive capability. The reference distribution is then defined as:
\begin{equation}
\mathbf{p}_{\text{ref}}=\operatorname{softmax}(\{-\ell(\mathbf{e}_k,\mathbf{e}_t)\}_{k=1}^K),
\end{equation}
assigning a higher probability to items that are more predictive of the target. The alignment loss aims to pull the attention distribution toward this utility-aware reference. Finally, the alignment loss is defined as the Kullback–Leibler divergence between the attention distribution $\mathbf{p}_{\text{aug}}$ and this utility-aware reference:
\begin{equation}
\label{eq:align}
\mathcal{L}_{\text{align}}=D_{KL}(\mathbf{p}_{\text{aug}}||\mathbf{p}_{\text{ref}}).
\end{equation}
Minimizing this term encourages the learned attention weights to not only reflect similarity but also align with each item's intrinsic contribution to predicting the target. In effect, the KL-divergence acts as a calibrator: it adjusts the similarity-based attention toward a distribution that prioritizes items with high predictive value, thereby ensuring that the aggregation step focuses on information that is truly beneficial for the final recommendation.

The overall training objective combines the recommendation loss and the alignment loss:
\begin{equation}
\label{eq:loss}
\mathcal{L}=\mathcal{L}_{\text{rec}}+\lambda \mathcal{L}_{\text{align}},
\end{equation}
where the $\lambda$ balances two terms. Notice that this design ensures the retrieval-learning module satisfies the following two complementary criteria:
\begin{enumerate}[label=(\arabic*), leftmargin=*, labelsep=0.5em, itemindent=0em,
                  topsep=0pt, noitemsep, nolistsep]
\item Preserving collaborative signals through the attention mechanism.
The cross-attention module naturally captures the similarity between the query (user sequence representation) and each retrieved item. This similarity reflects historical co-occurrence and interaction patterns encoded in the training data—i.e., the collaborative signals.
\item Emphasizing predictive utility via the KL-driven alignment with $\mathbf{p}_{\text{ref}}$. While similarity provides a useful prior, it does not guarantee that an item will help refine the current prediction. The KL-divergence term introduces a helpful objective: it aligns the attention distribution with a reference distribution that directly measures each retrieved item's ability to individually predict the target. This alignment effectively re-weights the retrieved items, amplifying those that are not only similar but also predictively discriminative for the specific target item. 
\end{enumerate}

Notice that the extra computation in our retrieval learning is marginal, requiring only the optimization of three matrices using the training data, and is model-agnostic, requiring no architectural assumptions about the underlying sequential recommendation model.

\subsection{Test-Time Retrieval Augmentation}
\label{sec:tta}

At test time, the SR model parameters remain frozen. The retrieval and fusion process operates entirely in forward pass, requiring no gradient computation or parameter updates. Once the augmented embedding $\mathbf{e}_{\text{aug}}$ is obtained, we first compute the corresponding augmentation-based prediction probability $p_{\text{aug}}$ via the same similarity computation used in Equation~\ref{eq:similarity}. Specifically:
\begin{equation}
\label{eq:pr}
p_{\text{aug}}(v = v^* \mid \mathbf{s}^{u_t}) = \frac{\exp\left(\mathbf{e}_{\text{aug}} \mathbf{e}_{v^*}^\top\right)}{\sum_{v_i \in \mathcal{V}} \exp\left(\mathbf{e}_{\text{aug}} \mathbf{e}_{v_i}^\top\right)}.
\end{equation}
With both the initial and augmentation predictions available, the next step is to refine the final prediction by effectively combining these two signals, as formulated in Equation~\ref{eq:refine}. We implement this refinement as follows:

\begin{equation}
\label{eq:refine_impl}
\begin{split}
p\bigl(v = v^* \mid \mathbf{s}^{u_t}\bigr)
&= \alpha \times p_{\text{init}}\bigl(v = v^* \mid \mathbf{s}^{u_t}\bigr) \\
&\quad + (1-\alpha) \, p_{\text{aug}}\bigl(v = v^* \mid \mathbf{s}^{u_t}\bigr),
\end{split}
\end{equation}
where $p_{\text{init}}$ is the initial prediction and $p_{\text{aug}}$ is the augmentation-based prediction, respectively. To adaptively determine the mixing coefficient $\alpha$, we propose an entropy-based dynamic fusion mechanism that weights each prediction based on its uncertainty across the entire item set.

\subsubsection{Uncertainty Estimation via Entropy}

We quantify the uncertainty of a prediction distribution by its entropy. For a probability vector $\mathbf{p}=[p_{v_1},\cdots,p_{v_{|\mathcal{V}|}}]$ obtained via softmax as Equation~\ref{eq:pr}, the entropy is computed as:
\begin{equation}
H(\mathbf{p}) = -\sum_{v_i\in\mathcal{V}} p_{v_i} \log(p_{v_i} + \epsilon),
\end{equation}
where $\epsilon = 10^{-8}$ ensures numerical stability.
Because the item set is typically large and follows a long-tail distribution, computing entropy over all items can dilute the discriminative signal. To focus on the most plausible candidates, we restrict the entropy calculation to the top-$\rho$ items with the highest logit scores:

\begin{equation}
\label{eq:fusion}
H_{\text{top}}(\mathbf{p}) = -\sum_{v_i \in \tau_{\text{top}}(\mathbf{p})} p_{v_i} \log\bigl(p_{v_i} + \epsilon\bigr),
\end{equation}
where $\tau_\text{top}(\mathbf{p})$ denotes the set of items whose logits rank in the top $\rho$ fraction. This truncation serves two purposes: (1) it concentrates the uncertainty measure on the region that matters most for ranking, and (2) it amplifies the relative difference in entropy between the two predictions, making the fusion weights more sensitive to genuine confidence variations.
\subsubsection{Confidence-Driven Fusion Weight}
Let $H_{\text{top}}^{\text{init}}$ and $H_{\text{top}}^{\text{aug}}$ be the truncated entropies of the initial and augmented predictions, respectively. the fusion weight $\alpha$ is computed as:
\begin{equation}
\alpha = \frac{\exp\left(\frac{1}{1+H^{\text{init}}_{\text{top}}}\right)}{\exp\left(\frac{1}{1+H^{\text{init}}_{\text{top}}}\right) + \exp\left(\frac{1}{1+H^{\text{aug}}_{\text{top}}}\right)}.
\end{equation}
The rationale behind this formulation is that lower entropy corresponds to higher prediction confidence. When a model outputs a sharply peaked distribution (low $H$), it expresses strong discriminative certainty among the top candidates, and should therefore receive a larger weight in the fusion. Conversely, a flat distribution (high $H$) reflects uncertainty and is assigned a smaller weight.

This entropy-guided weighting scheme enables ReAd to dynamically balance the contributions of the original sequential representation and the retrieval-augmented signal, seamlessly adapting to the varying confidence levels across different user sequences at test time.

%% file: section/experiment.tex
\section{Experiments}
\label{sec:exp}
In this section, we conduct extensive experiments to address the following research questions:

\begin{itemize}[leftmargin=*,topsep=0pt,noitemsep,nolistsep]
\item \textbf{RQ1}: Does ReAd consistently outperform existing sequential recommendation baselines?
\item \textbf{RQ2}: How well does ReAd generalize across different backbone SR architectures?
\item \textbf{RQ3}: How do key hyperparameters in the retrieval learning module, i.e., the number of retrieved items $K$ and the fusion weight $\lambda$, affect the performance of ReAd?
\item \textbf{RQ4}: What is the contribution of each core component in ReAd, and is the introduced computational overhead acceptable for real-time inference?
\item \textbf{RQ5}: Does ReAd indeed retrieve semantically relevant and helpful items during test-time adaptation?
\end{itemize}

\subsection{Experimental Settings}
\subsubsection{Datasets.}
We evaluate our method using five public datasets that represent diverse recommendation scenarios, which are widely adopted in previous sequential recommendation research. Four Amazon subsets—Office, Beauty, Sports, and Home—span distinct product categories and exhibit varied interaction sparsity, reflecting real-world e-commerce environments. The ML-1M dataset provides a widely adopted benchmark in the movie domain with denser user activity.
Following common practice in sequential recommendation~\cite{tta,duorec}, we filter out users and items with fewer than five interactions to ensure data quality and mitigate extreme sparsity. Table~\ref{tab:dataset_stats} summarizes the resulting dataset statistics.

\begin{table}[!htbp]
    \centering
    \small
    \setlength{\tabcolsep}{6pt}
    \renewcommand{\arraystretch}{0.90}
    \caption{Statistics of the datasets used in experiments.}
    \begin{tabular}{l c c c c c}
        \toprule
        Statistic       & Office  & Beauty  & Sports  & Home    & ML-1M   \\
        \midrule
        \#Users         & 4,906   & 22,364  & 35,599  & 66,520  & 6,040   \\
        \#Items         & 2,421   & 12,102  & 18,358  & 28,238  & 3,706   \\
        \#Interactions  & 53,258  & 198,502 & 296,337 & 551,682 & 1,000,209 \\
        Avg. Actions    & 10.86   & 8.88    & 8.32    & 8.29    & 165.60  \\
        Sparsity        & 99.55\% & 99.93\% & 99.95\% & 99.97\% & 95.53\% \\
        \bottomrule
        \end{tabular}
    \label{tab:dataset_stats}
\end{table}

\subsubsection{Baselines.} 

To conduct a comprehensive evaluation, we include twelve baselines that fall into three categories as outlined in Section~\ref{sec:rw}. The first group comprises architectural models designed to capture sequential patterns, including \textbf{GRU4Rec}~\cite{GRU4Rec}, \textbf{SASRec}~\cite{sasrec}, and \textbf{BERT4Rec}~\cite{bert4rec}.

The second group consists of contrastive learning-based methods that address data sparsity through augmentation and alignment. \textbf{CL4SRec}~\cite{cl4srec} and \textbf{DuoRec}~\cite{duorec} use sequence augmentation strategies. \textbf{CoSeRec}~\cite{coserec} and \textbf{MCLRec}~\cite{mcl} further improve augmentation and alignment. \textbf{S$^3$-Rec}~\cite{s3}, \textbf{ICLRec}~\cite{ICL}, and \textbf{ICSRec}~\cite{ics} instead align different sequence views to improve representation learning.

The third group covers test-time and retrieval-augmented enhancement methods: \textbf{TTA}~\cite{tta} applies random perturbations at inference time, while \textbf{RaSeRec}~\cite{Raserec} incorporates retrieval during training but is included as a strong retrieval-based baseline.

\subsubsection{Evaluation Metrics.} To ensure an unbiased evaluation, we adopt the leave-one-out strategy to split each user’s interaction sequence into training, validation, and test segments. During evaluation, we rank all items in the catalog for every test sequence and compute metrics over the full ranking list, avoiding the potential bias introduced by negative-item sampling. We employ two widely used ranking-based metrics: Hit Ratio@K (HR@K) and Normalized Discounted Cumulative Gain@K (ND@K) with $K\in\{5, 10, 20 \}$. Higher values of both metrics indicate better ranking performance. 

\subsubsection{Implementation Details.} 
For all baseline models, we use the open-source implementations provided in RecBole~\cite{recbole} under their recommended settings. Our proposed ReAd framework is intentionally model-agnostic. To demonstrate this property, we instantiate it on representative models from both the architectural group (SASRec) and the contrastive-learning group (DuoRec), thereby showing that ReAd can be flexibly combined with different types of sequential recommendation backbones.

We set the embedding dimension to 64 and the batch size to 256 across all experiments. All other hyperparameters for the baselines are carefully tuned following the configurations described in their original papers. Training is performed with the Adam optimizer~\cite{adam} using a learning rate of $0.001$
The three related hyperparameters in our method are the $K$ values in Equation~\ref{eq:topk} and $\lambda$ in Equation~\ref{eq:loss}, which balance the recommendation and alignment objectives. The top ratio in Equation~\ref{eq:fusion}, which controls the fraction of items considered in the entropy-based fusion. We hence search $K \in \{1,3,5,10,15,20\}$, $\lambda\in\{0,0.5,1,1.5,2\}$, and top ratio in $\rho\in\{10\%,5\%,1\%,0.5\%,0.1\%\}$. All experiments are conducted on an NVIDIA GeForce RTX 4090D and run for ten runs, reporting the average results for all methods.

\begin{table*}[!htbp]
    \centering
    \small
    \setlength{\tabcolsep}{4pt}
    \renewcommand{\arraystretch}{0.78}
    \caption{Performance on five public datasets. Bold and underlined values denote the best and second-best results; ReAd's gains are significant under paired t-tests ($p<0.01$).}
    \resizebox{\textwidth}{!}{%
    \begin{tabular}{l l | c c c | c c c c c c c | c c | c c}
        \toprule
        Dataset & Metric                    & GRU4Rec& SASRec &BERT4Rec& S³-Rec & CL4SRec& CoSeRec& ICLRec & DuoRec & MCLRec & ICSRec & RaSeRec            & TTA    & \makecell[c]{ReAd\\(+SASRec)} & \makecell[c]{ReAd\\(+DuoRec)}     \\
        \midrule
        \multirow{6}{*}{Office}   & HR@5    & 0.0277 & 0.0544 & 0.0376 & 0.0443 & 0.0480 & 0.0577 & 0.0564 & 0.0644 & \underline{0.0671} & 0.0651 & 0.0669 & 0.0548 & 0.0585 & \textbf{0.0698} \\
                                  & HR@10   & 0.0532 & 0.0899 & 0.0666 & 0.0658 & 0.0665 & 0.0811 & 0.0815 & 0.1011 & \underline{0.1071} & 0.1051 & 0.1042 & 0.0896 & 0.0907 & \textbf{0.1090} \\
                                  & HR@20   & 0.0985 & 0.1388 & 0.1166 & 0.1221 & 0.1121 & 0.1346 & 0.1195 & 0.1552 & \underline{0.1648} & 0.1576 & 0.1641 & 0.1389 & 0.1427 & \textbf{0.1668} \\
                                  & ND@5    & 0.0169 & 0.0369 & 0.0233 & 0.0296 & 0.0308 & 0.0394 & 0.0345 & 0.0410 & 0.0407 & 0.0416 & \underline{0.0429} & 0.0371  & 0.0384 & \textbf{0.0456}\\
                                  & ND@10   & 0.0252 & 0.0483 & 0.0326 & 0.0383 & 0.0385 & 0.0500 & 0.0439 & 0.0527 & 0.0538 & 0.0525 & \underline{0.0549} & 0.0492  & 0.0489 & \textbf{0.0582}\\
                                  & ND@20   & 0.0365 & 0.0605 & 0.0452 & 0.0523 & 0.0502 & 0.0611 & 0.0534 & 0.0663 & 0.0691 & 0.0679 & \underline{0.0699} & 0.0625  & 0.0620 & \textbf{0.0727}\\
        \midrule
        \multirow{6}{*}{Beauty}   & HR@5    & 0.0206 & 0.0377 & 0.0340 & 0.0395 & 0.0471 & 0.0506 & 0.0498 & 0.0538 & 0.0563 & 0.0540 & \underline{0.0570} & 0.0416  & 0.0447 & \textbf{0.0582}\\
                                  & HR@10   & 0.0325 & 0.0598 & 0.0526 & 0.0619 & 0.0654 & 0.0726 & 0.0741 & 0.0825 & \underline{0.0869} & 0.0841 & 0.0865 & 0.0629  & 0.0687 & \textbf{0.0874}\\
                                  & HR@20   & 0.0499 & 0.0877 & 0.0789 & 0.0937 & 0.0985 & 0.1031 & 0.1059 & 0.1184 & 0.1212 & 0.1198 & \underline{0.1221} & 0.0921  & 0.0967 & \textbf{0.1243}\\
                                  & ND@5    & 0.0127 & 0.0237 & 0.0222 & 0.0251 & 0.0273 & 0.0339 & 0.0331 & 0.0340 & 0.0344 & 0.0338 & \underline{0.0369} & 0.0266 & 0.0290 & \textbf{0.0386} \\
                                  & ND@10   & 0.0166 & 0.0308 & 0.0282 & 0.0323 & 0.0350 & 0.0410 & 0.0405 & 0.0433 & 0.0446 & 0.0435 & \underline{0.0461} & 0.0329  & 0.0367 & \textbf{0.0481}\\
                                  & ND@20   & 0.0209 & 0.0378 & 0.0349 & 0.0403 & 0.0410 & 0.0488 & 0.0488 & 0.0523 & 0.0539 & 0.0525 & \underline{0.0554} & 0.0419 & 0.0438 & \textbf{0.0573} \\
        \midrule
        \multirow{6}{*}{Sports}   & HR@5    & 0.0107 & 0.0216 & 0.0170 & 0.0220 & 0.0256 & 0.0285 & 0.0291 & 0.0310 & \underline{0.0322} & 0.0316 & 0.0315 & 0.0239  & 0.0233 & \textbf{0.0331}\\
                                  & HR@10   & 0.0178 & 0.0326 & 0.0281 & 0.0336 & 0.0382 & 0.0426 & 0.0429 & 0.0471 & \underline{0.0486} & 0.0479 & 0.0487 & 0.0398  & 0.0339 & \textbf{0.0496}\\
                                  & HR@20   & 0.0279 & 0.0479 & 0.0444 & 0.0510 & 0.0553 & 0.0636 & 0.0639 & 0.0692 & \underline{0.0720} & 0.0712 & 0.0711 & 0.0637  & 0.0500 & \textbf{0.0726}\\
                                  & ND@5    & 0.0068 & 0.0148 & 0.0109 & 0.0147 & 0.0150 & 0.0179 & 0.0181 & 0.0193 & \underline{0.0204} & 0.0202 & 0.0196 & 0.0168  & 0.0163 & \textbf{0.0221}\\
                                  & ND@10   & 0.0091 & 0.0184 & 0.0144 & 0.0185 & 0.0217 & 0.0231 & 0.0238 & 0.0245 & \underline{0.0260} & 0.0245 & 0.0251 & 0.0214  & 0.0197 & \textbf{0.0274}\\
                                  & ND@20   & 0.0116 & 0.0222 & 0.0185 & 0.0229 & 0.0244 & 0.0275 & 0.0286 & 0.0300 & \underline{0.0311} & 0.0301 & 0.0308 & 0.0266  & 0.0237 & \textbf{0.0332} \\
        \midrule
        \multirow{6}{*}{Home}     & HR@5    & 0.0055 & 0.0081 & 0.0083 & 0.0103 & 0.0136 & 0.0153 & 0.0146 & 0.0190 & \underline{0.0198} & 0.0198 & 0.0196 & 0.0108 & 0.0105 & \textbf{0.0203}\\
                                  & HR@10   & 0.0104 & 0.0131 & 0.0143 & 0.0155 & 0.0198 & 0.0232 & 0.0225 & 0.0278 & 0.0286 & 0.0289 & \underline{0.0293} & 0.0175  & 0.0164 & \textbf{0.0298}\\
                                  & HR@20   & 0.0180 & 0.0204 & 0.0241 & 0.0260 & 0.0282 & 0.0336 & 0.0345 & 0.0402 & \underline{0.0419} & 0.0418 & 0.0409 & 0.0266  & 0.0248 & \textbf{0.0426}\\
                                  & ND@5    & 0.0034 & 0.0052 & 0.0052 & 0.0074 & 0.0079 & 0.0107 & 0.0101 & 0.0117 & 0.0122 & \underline{0.0128} & 0.0121 & 0.0069 & 0.0071 & \textbf{0.0133} \\
                                  & ND@10   & 0.0049 & 0.0068 & 0.0072 & 0.0092 & 0.0100 & 0.0133 & 0.0133 & 0.0145 & 0.0150 & \underline{0.0152} & 0.0152 & 0.0091 & 0.0090 & \textbf{0.0163} \\
                                  & ND@20   & 0.0068 & 0.0086 & 0.0096 & 0.0115 & 0.0121 & 0.0159 & 0.0161 & 0.0176 & \underline{0.0185} & 0.0179 & 0.0181 & 0.0113  & 0.0111 & \textbf{0.0196}\\
        \midrule
        \multirow{6}{*}{ML-1M}    & HR@5    & 0.0462 & 0.1364 & 0.1364 & 0.1192 & 0.1163 & 0.1117 & 0.1312 & 0.1909 & 0.1889 & 0.1913 & \underline{0.1932} & 0.1359 & 0.1644 & \textbf{0.1956} \\
                                  & HR@10   & 0.0654 & 0.2119 & 0.2156 & 0.2079 & 0.2006 & 0.1878 & 0.2196 & \underline{0.2859} & 0.2832 & 0.2799 & 0.2844 & 0.2136  & 0.2382 & \textbf{0.2897}\\
                                  & HR@20   & 0.0980 & 0.3144 & 0.3164 & 0.3181 & 0.3121 & 0.2997 & 0.3357 & 0.3839 & 0.3796 & 0.3844 & \underline{0.3879} & 0.3175 & 0.3404 & \textbf{0.3897} \\
                                  & ND@5    & 0.0299 & 0.0894 & 0.0902 & 0.0748 & 0.0738 & 0.0693 & 0.0844 & 0.1297 & 0.1288 & 0.1287 & \underline{0.1328} & 0.0814 & 0.1080 & \textbf{0.1341} \\
                                  & ND@10   & 0.0360 & 0.1135 & 0.1156 & 0.1120 & 0.0989 & 0.0981 & 0.1116 & 0.1603 & 0.1600 & 0.1588 & \underline{0.1646} & 0.1105  & 0.1317 & \textbf{0.1653}\\
                                  & ND@20   & 0.0442 & 0.1393 & 0.1410 & 0.1347 & 0.1289 & 0.1269 & 0.1371 & 0.1850 & 0.1833 & 0.1851 & \underline{0.1888} & 0.1372 & 0.1575 & \textbf{0.1907} \\
        \bottomrule
        \end{tabular}%
    }
    \label{tab:recsys_main}
\end{table*}
\subsection{Overall Performance (RQ1)}

We compare our ReAd with the mentioned baselines, and the overall results are illustrated in Table~\ref{tab:recsys_main}. We make several observations as follows. 

\begin{itemize}[leftmargin=*,topsep=0pt,noitemsep,nolistsep]

\item
When built upon DuoRec as the backbone, ReAd consistently achieves the best results across all five datasets, demonstrating its effectiveness. The performance gap between ReAd (+DuoRec) and both ReAd (+SASRec) and the test-time baselines (TTA and RaSeRec) also highlights the benefit of incorporating contrastive learning during training. Notably, RaSeRec—which introduces retrieval through a dedicated pre-training architecture but does not use contrastive learning—performs comparably to TTA, while ReAd (+DuoRec) outperforms both. This confirms that ReAd is model-agnostic and can flexibly leverage different training paradigms (e.g., contrastive learning) to further boost performance. 

\item
Compared with the original SASRec, both TTA and ReAd (+SASRec) improve the performance, confirming the value of test-time augmentation. However, ReAd (+SASRec) consistently surpasses TTA, indicating that our retrieval-augmented refinement, which integrates both retrieval-based augmentation and confidence-aware prediction fusion, is more effective than TTA’s purely input-level augmentation.

\item This pattern suggests that test-time retrieval augmentation is particularly beneficial when historical signals are sparse, where the frozen model alone is insufficient. Both RaSeRec and ReAd, which employ retrieval mechanisms, show greater improvements on sparse data, reinforcing the importance of augmenting sparse sequences with externally retrieved collaborative signals.
\end{itemize}

\subsection{Generalize to Other Architectures (RQ2)}

\begin{table}[!htbp]
    \centering
    \small
    \setlength{\tabcolsep}{4pt}
    \renewcommand{\arraystretch}{0.90}
    \caption{ReAd improvements on different backbones (all $p<0.01$).}
    \begin{tabular}{l l | c c c | c c c }
        \toprule
        \multirow{2}{*}{Dataset} & \multirow{2}{*}{Metric} & \multicolumn{3}{c|}{GRU4Rec} & \multicolumn{3}{c}{BERT4Rec} \\
         &  & Orig & ReAd & Imp. & Orig & ReAd & Imp.  \\
        \midrule
        \multirow{4}{*}{Office}  & HR@5  & 0.0277 & 0.0320 & 15.52\% & 0.0376 & 0.0437 & 16.22\% \\
                                 & HR@10 & 0.0532 & 0.0595 & 11.84\% & 0.0666 & 0.0742 & 11.41\%  \\
                                 & ND@5  & 0.0169 & 0.0199 & 17.75\% & 0.0233 & 0.0271 & 16.31\%  \\
                                 & ND@10 & 0.0252 & 0.0288 & 14.29\% & 0.0326 & 0.0369 & 13.19\%  \\
        \midrule
        \multirow{4}{*}{Beauty}  & HR@5  & 0.0206 & 0.0225 & 9.22\%  & 0.0340 & 0.0372 & 9.41\% \\
                                 & HR@10 & 0.0325 & 0.0344 & 5.85\%  & 0.0526 & 0.0585 & 11.22\% \\
                                 & ND@5  & 0.0127 & 0.0147 & 15.75\% & 0.0222 & 0.0242 & 9.01\%  \\
                                 & ND@10 & 0.0166 & 0.0186 & 12.05\% & 0.0282 & 0.0311 & 10.28\%  \\
        \midrule
        \multirow{4}{*}{Sports}  & HR@5  & 0.0107 & 0.0119 & 11.21\% & 0.0170 & 0.0190 & 11.76\% \\
                                 & HR@10 & 0.0178 & 0.0189 & 6.18\%  & 0.0281 & 0.0305 & 8.54\%  \\
                                 & ND@5  & 0.0068 & 0.0078 & 14.71\% & 0.0109 & 0.0124 & 13.76\% \\
                                 & ND@10 & 0.0091 & 0.0100 & 9.89\%  & 0.0144 & 0.0161 & 11.81\% \\
        \midrule
        \multirow{4}{*}{Home}    & HR@5  & 0.0055 & 0.0059 & 7.27\%  & 0.0083 & 0.0093 & 12.05\% \\
                                 & HR@10 & 0.0104 & 0.0109 & 4.81\%  & 0.0143 & 0.0159 & 11.19\%  \\
                                 & ND@5  & 0.0034 & 0.0036 & 5.88\%  & 0.0052 & 0.0058 & 11.54\%  \\
                                 & ND@10 & 0.0049 & 0.0052 & 6.12\%  & 0.0072 & 0.0079 & 9.72\% \\
        \midrule
        \multirow{4}{*}{ML-1M}   & HR@5  & 0.0462 & 0.0505 & 9.31\%  & 0.1364 & 0.1412 & 3.52\%  \\
                                 & HR@10 & 0.0654 & 0.0733 & 12.08\% & 0.2156 & 0.2243 & 4.04\% \\
                                 & ND@5  & 0.0299 & 0.0338 & 13.04\% & 0.0902 & 0.0929 & 2.99\%  \\
                                 & ND@10 & 0.0360 & 0.0409 & 13.61\% & 0.1156 & 0.1196 & 3.46\%  \\
        \bottomrule
        \end{tabular}%
        
\label{tab:tta_backbone_improve}
\end{table}

To further verify the generalization capability of ReAd, we apply it to SR models with different architectural designs: SASRec and DuoRec use causal (unidirectional) self-attention; BERT4Rec employs regular (bidirectional) self-attention; and GRU4Rec relies on gated recurrent units~\cite{reviewforbert4rec}. The results are summarized in Table~\ref{tab:tta_backbone_improve}, along with the earlier SASRec results in Table~\ref{tab:recsys_main}. 

ReAd consistently improves recommendation performance across all tested architectures, confirming its model-agnostic design. Moreover, the magnitude of improvement varies, with most gains exceeding 10\%, although a few remain modest. This is attributed to the inherent capacity of the backbone model: stronger base models may leave less room for relative improvement. Additionally, dataset characteristics play a role: the gains are larger on datasets where sequential patterns are sparse or exhibit greater diversity, as test-time retrieval augmentation provides complementary signals that the frozen model lacks. Overall, these experiments demonstrate that ReAd effectively enhances a diverse set of sequential recommendation architectures, validating its robustness and practical applicability. 

\subsection{Hyperparameter Sensitivity (RQ3)}

\begin{figure}[!htbp]
	\begin{minipage}[t]{\linewidth}
		\centering
		\includegraphics[width=0.82\linewidth]{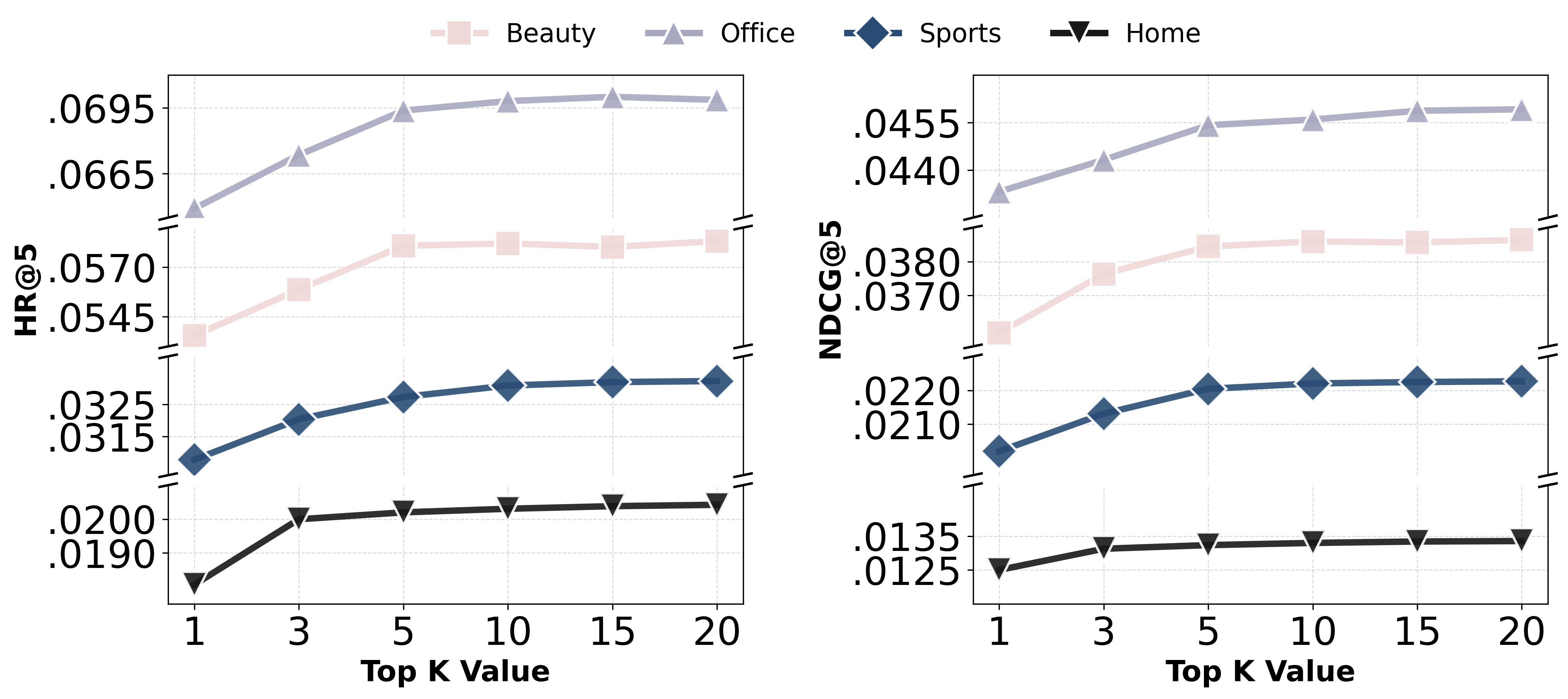}
	\end{minipage}
	\begin{minipage}[t]{\linewidth}
		\centering
		\includegraphics[width=0.82\linewidth]{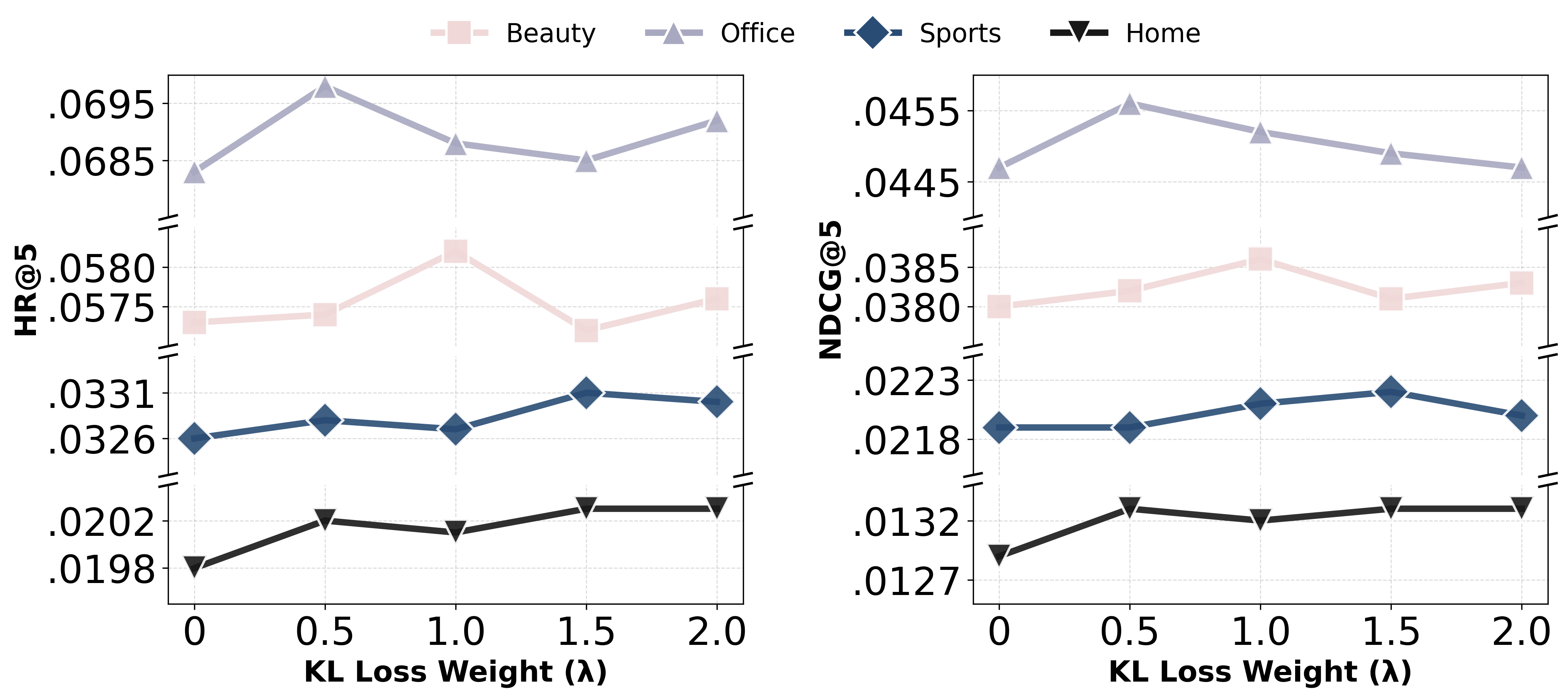}
		\end{minipage}
		\caption{Effects of the number of retrieved items $K$ and KL-alignment coefficient $\lambda$.}
		\Description{Two line-chart panels report recommendation performance across datasets as the number of retrieved items $K$ and the alignment-loss weight $\lambda$ vary.}
		\label{fig:para}
\end{figure}

As discussed in Section~\ref{sec:method}, ReAd involves three key hyperparameters. The first two are the number of retrieved items $K$ and the alignment-loss coefficient $\lambda$ in Equation~\ref{eq:loss}, which govern retrieval learning. As shown in Figure~\ref{fig:para} (top), increasing $K$ initially improves performance, peaking at $K=10$ for most datasets, after which performance gradually declines. This confirms that retrieving too many items introduces irrelevant or noisy signals. The bottom subfigure shows that varying $\lambda$ has a relatively modest impact, suggesting that the alignment loss plays a supportive role when retrieved items are already collaboratively relevant.

Across different backbone models, performance follows a consistent trend: it initially improves with larger $K$, peaks at an optimal value, and then declines. This pattern can be explained as follows: a moderate increase in $K$ enriches the augmentation signal with more collaborative information, but beyond a certain point, irrelevant or noisy items are introduced, which hinder the learning process. Therefore, selecting an appropriate $K$ is crucial for achieving the best improvement. In contrast, varying $\lambda$ does not lead to significant changes in performance. We hypothesize that this is because the retrieved set is relatively small and the items are already collaboratively relevant, so strictly aligning the attention distribution with the reference distribution yields diminishing returns.

\begin{figure}[!htbp]
    \centering
    \includegraphics[width=0.9\linewidth]{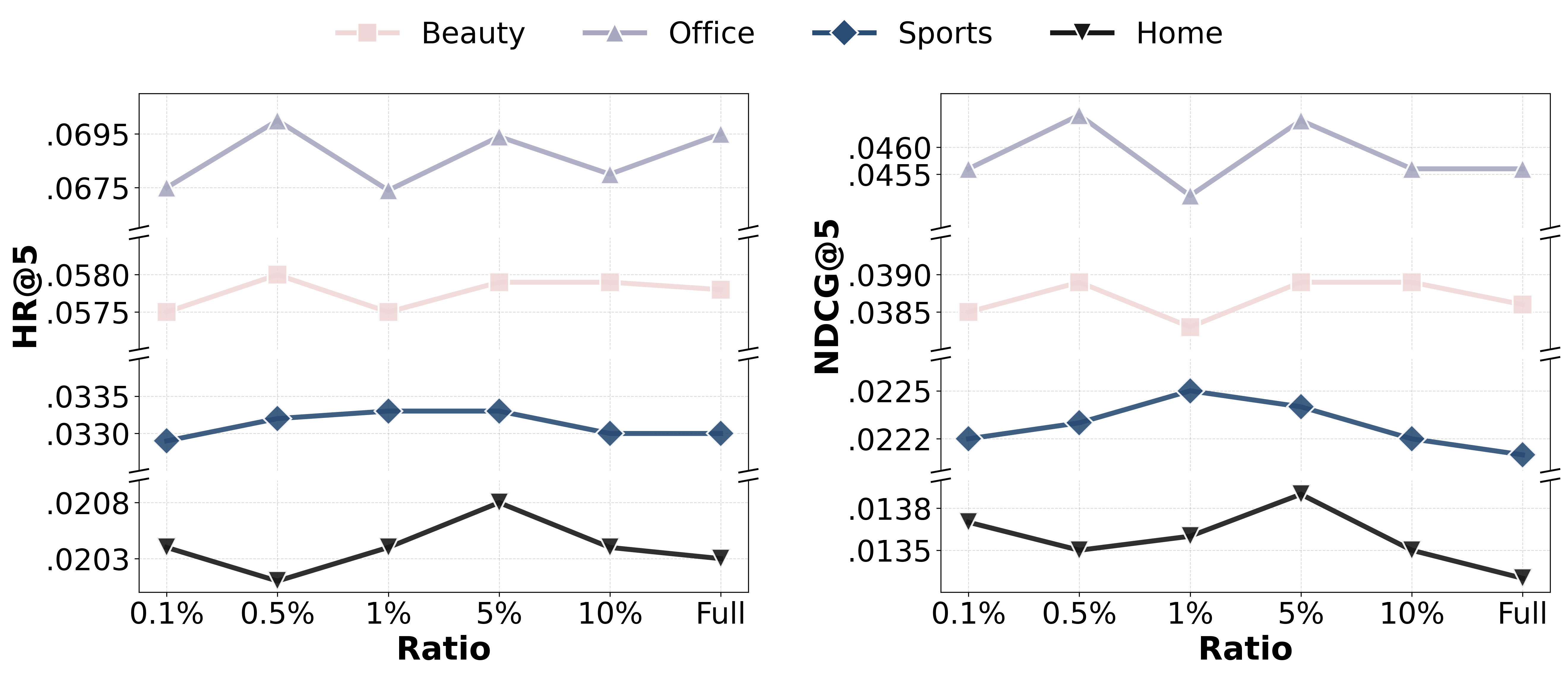}
    \caption{Effect of top-ratio $\rho$ in $\tau_{\mathrm{top}}$.}
    \Description{A line chart compares recommendation performance across datasets for different top-ratio values used in the entropy calculation.}
    \label{fig:change}
\end{figure}

The fraction $\rho$ of top-ranked items considered in $\tau_{top}$ directly influences the entropy-based fusion weight $\alpha$ in Equation~\ref{eq:refine}, and consequently the final refined prediction. As shown in Figure~\ref{fig:change}, a larger top-ratio generally leads to better performance, whereas using the full item set (i.e., top-ratio $= 100\%$) consistently degrades results, though more stable. This can be explained by the long-tail nature of item catalogs: when entropy is computed over all items, the numerous low-probability tail items dilute the discriminative signal, making the entropies of the initial and augmented predictions too similar.

%The number of top ratio items in $\tau_{top}$ affects the test time adaptation via refining the final prediction in Equation~\ref{eq:refine}. From Figure~\ref{fig:change}, we can observe that an increase in the number of top ratio items leads to an improvement in performance in general. On the other hand, the $\alpha$ based on the full item, despite being more stable, will degrade performance. The reason is primarily that the long-tailed items make the entropy of two predictions more similar, and calculating $\alpha$ on the top items will thus be more helpful for our method to distinguish the top-ranked items.

\subsection{Ablation Study (RQ4)}

\begin{table}[!htbp]
    \centering
    \renewcommand{\arraystretch}{0.90}
    \caption{Ablation results on four Amazon datasets with DuoRec.}
    \label{tab:ablation}
    \resizebox{\linewidth}{!}{
    \begin{tabular}{l|cccc|cccc}
        \toprule
        \multirow{2}{*}{Method} & \multicolumn{4}{c|}{HR@10} & \multicolumn{4}{c}{NDCG@10} \\
        \cmidrule(lr){2-5} \cmidrule(lr){6-9}
         & Beauty & Home & Office & Sports & Beauty & Home & Office & Sports \\
        \midrule
        w/o Att & 0.0786 & 0.0268 & 0.0901 & 0.0436 & 0.0413 & 0.0127 & 0.0484 & 0.0207 \\
        w/o KL & 0.0863 & 0.0293 & 0.1074 & 0.0489 & 0.0473 & 0.0159 & 0.0571 & 0.0269 \\
        w/o Rec & 0.0643 & 0.0218 & 0.0752 & 0.0345 & 0.0332 & 0.0095 & 0.0381 & 0.0198 \\
        w/o $\alpha$ & 0.0845 & 0.0279 & 0.1044 & 0.0479 & 0.0451 & 0.0149 & 0.0536 & 0.0243 \\
        \midrule
        \textbf{ReAd} & \textbf{0.0874} & \textbf{0.0298} & \textbf{0.1090} & \textbf{0.0496} & \textbf{0.0481} & \textbf{0.0163} & \textbf{0.0582} & \textbf{0.0274} \\
        \bottomrule
    \end{tabular}
    }
\end{table}

In this section, we conduct ablation studies to investigate the contribution of each component in ReAd. All results are reported with DuoRec as the backbone (i.e., DuoRec+ReAd) on four Amazon datasets. The variants and their results are summarized in Table~\ref{tab:ablation}:

\begin{itemize}[leftmargin=*,topsep=0pt,noitemsep,nolistsep]
\item \textbf{w/o Rec}: Removes $\mathcal{L}_{\text{rec}}$. The retrieval module is trained only with $\mathcal{L}_{\text{align}}$, without direct supervision to predict the target item.

\item \textbf{w/o KL}: Removes $\mathcal{L}_{\text{align}}$. The retrieval module is trained only with $\mathcal{L}_{\text{rec}}$, without calibration against predictive utility.

\item \textbf{w/o Att}: Replaces learnable cross-attention with simple averaging of retrieved embeddings.

\item \textbf{w/o $\alpha$}: Fixes $\alpha = 0.5$, disabling entropy-based dynamic fusion.
\end{itemize}

Based on the results, we make the following observations. First, \textit{w/o Rec} causes the largest drop, confirming that retrieval must be guided by the prediction objective. Second, \textit{w/o Att} underperforms, showing that simple averaging fails to capture the varying predictive utility of retrieved items. Third, \textit{w/o $\alpha$} also harms performance, underscoring the importance of dynamic fusion. Finally, \textit{w/o KL} shows the smallest drop, as retrieved items are already relevant via similarity.

To directly address the concern of whether retrieval is necessary, we compare ReAd against two ablation strategies: (i) \textbf{Random Retrieval}, which randomly samples $K$ items from the item set; and (ii) \textbf{Popularity Retrieval}, which always retrieves the $K$ most popular items in the training set. Table~\ref{tab:retrieval_ablation} summarizes the results.
Several observations can be given. First, random retrieval consistently degrades performance across all datasets, confirming that indiscriminate retrieval introduces noise. Second, popularity retrieval shows mixed results: it underperforms ReAd on Beauty and Office, but slightly outperforms on the extremely sparse Home dataset, where popularity bias provides a simple but effective alternative when similarity-based retrieval struggles due to data sparsity. Third, ReAd achieves the best or comparable results across all datasets, demonstrating that similarity-based retrieval is the most robust strategy for test-time augmentation.

\begin{table}[!t]
    \centering
    \small
    \setlength{\tabcolsep}{4pt}
    \renewcommand{\arraystretch}{0.86}
    \caption{Retrieval-strategy comparison on four Amazon datasets (best in bold).}
    \begin{tabular}{l l c c c c}
        \toprule
        Dataset & Metric & SASRec & Random & Popularity & \textbf{ReAd} \\
        \midrule
        \multirow{2}{*}{Beauty} & NDCG@10 & 0.0308 & 0.0212 & 0.0226 & \textbf{0.0367} \\
        & HR@10 & 0.0598 & 0.0421 & 0.0453 & \textbf{0.0687} \\
        \midrule
        \multirow{2}{*}{Office} & NDCG@10 & 0.0483 & 0.0360 & 0.0378 & \textbf{0.0489} \\
        & HR@10 & 0.0899 & 0.0734 & 0.0744 & \textbf{0.0907} \\
        \midrule
        \multirow{2}{*}{Sports} & NDCG@10 & 0.0184 & 0.0128 & 0.0196 & \textbf{0.0197} \\
        & HR@10 & 0.0326 & 0.0224 & 0.0321 & \textbf{0.0339} \\
        \midrule
        \multirow{2}{*}{Home} & NDCG@10 & 0.0068 & 0.0050 & 0.0090 & 0.0090 \\
        & HR@10 & 0.0131 & 0.0100 & 0.0163 & \textbf{0.0164} \\
        \bottomrule
    \end{tabular}
    \label{tab:retrieval_ablation}
\end{table}

\begin{figure*}[!t]
    \centering
    \begin{tabular}{@{}c@{\hspace{2pt}}c@{\hspace{2pt}}c@{\hspace{2pt}}c@{}}
        \includegraphics[width=0.22\textwidth]{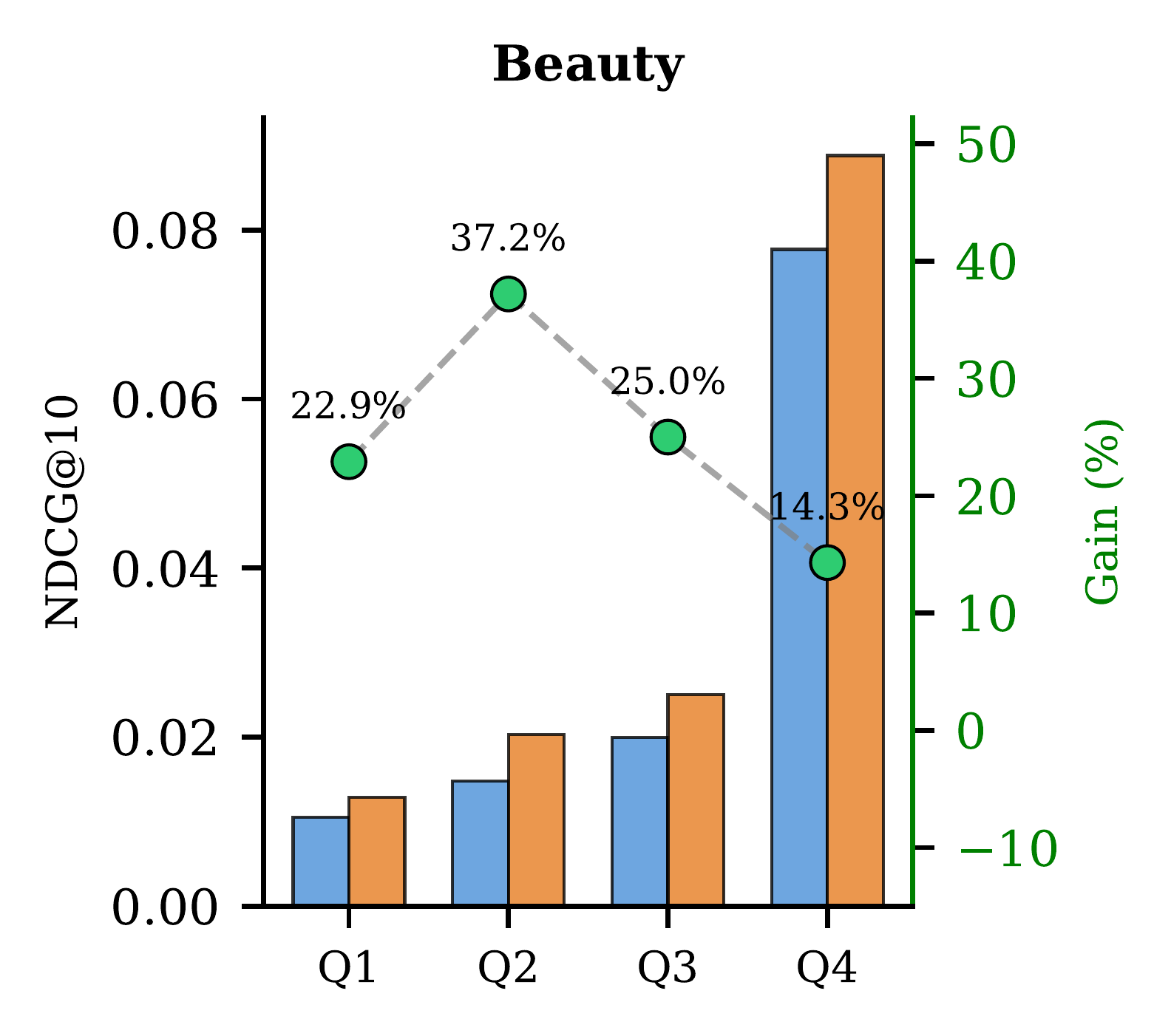} &
        \includegraphics[width=0.22\textwidth]{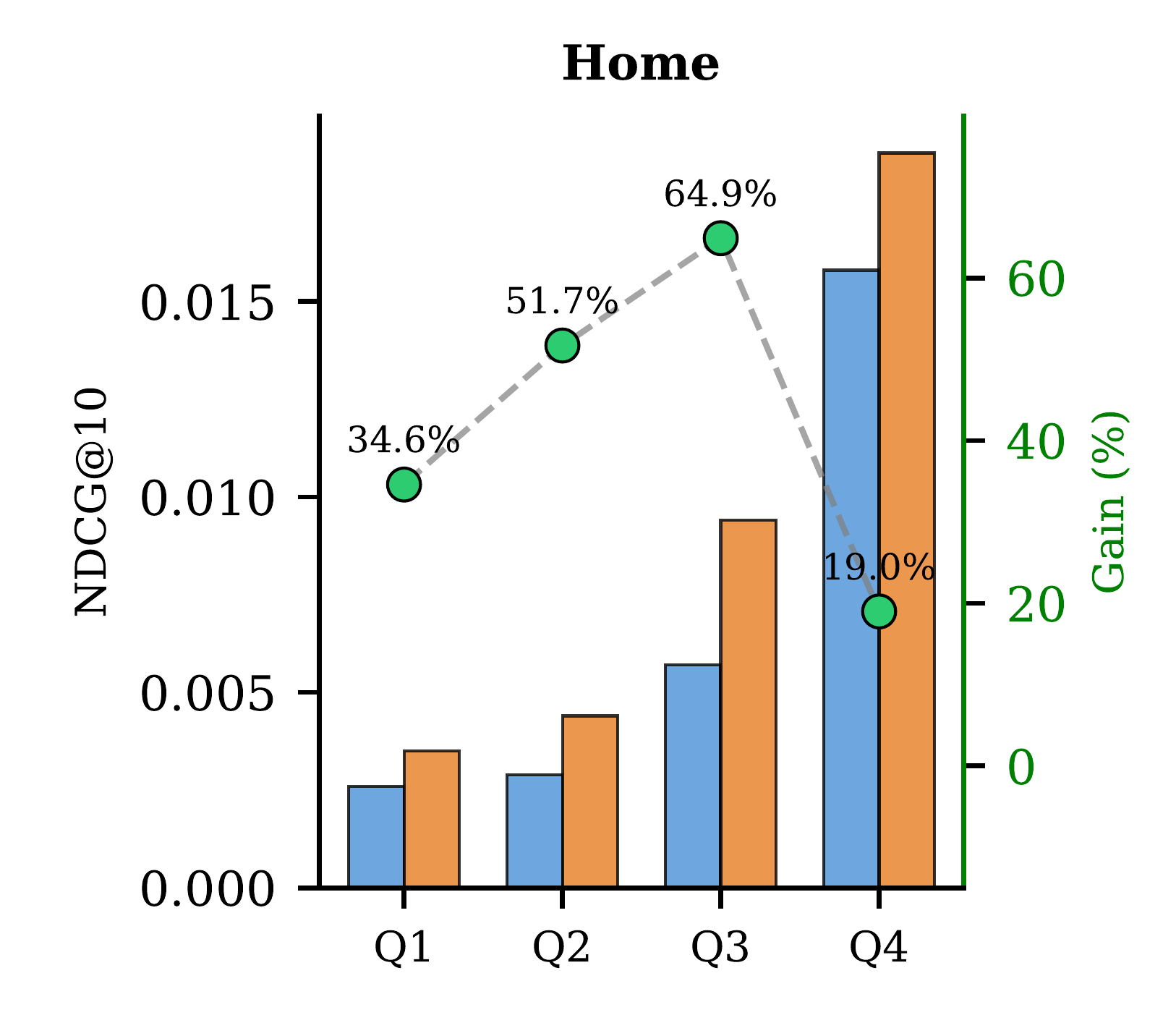} &
        \includegraphics[width=0.22\textwidth]{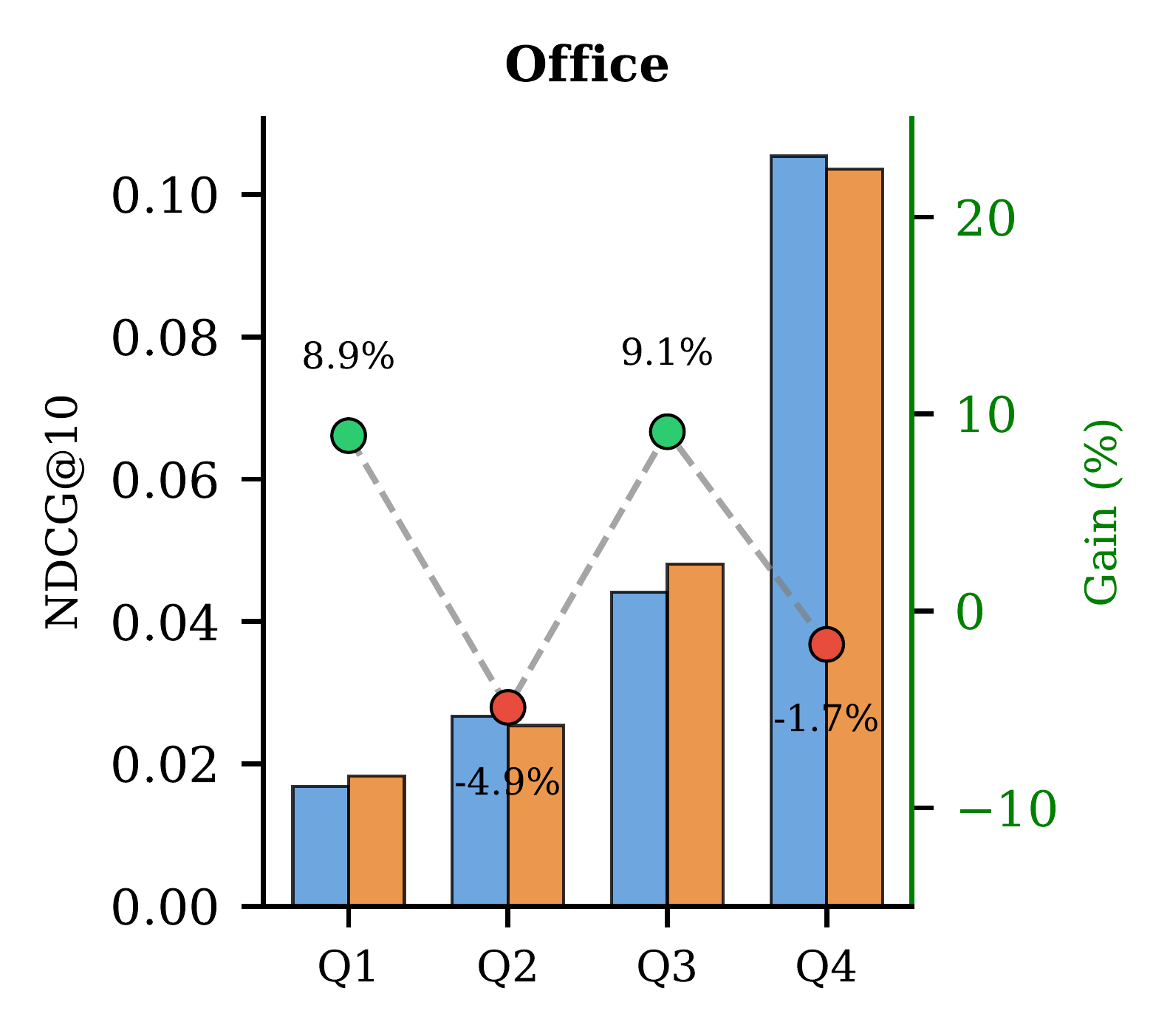} &
        \includegraphics[width=0.22\textwidth]{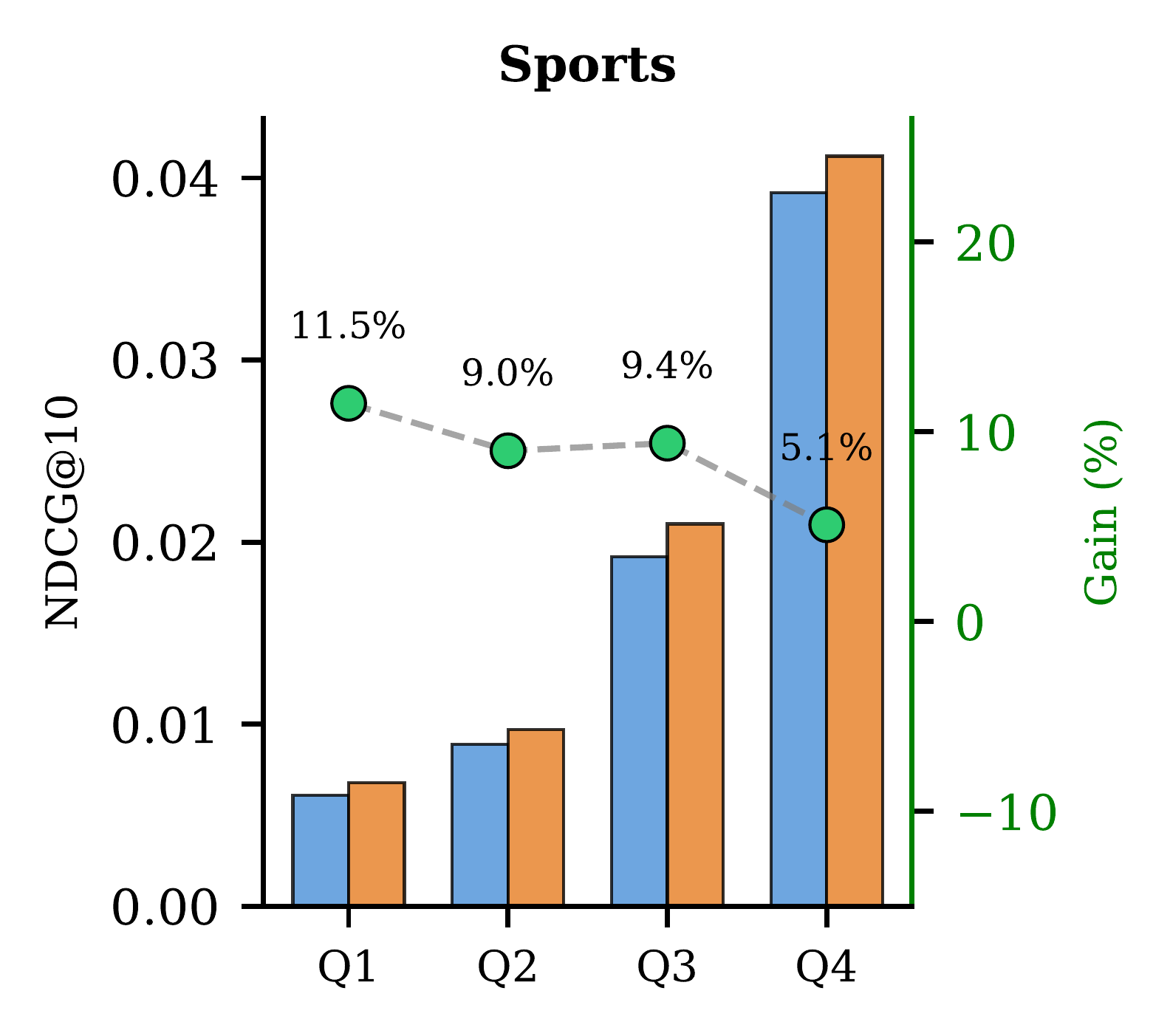} \\
    \end{tabular}
    \caption{Item popularity stratified results on four Amazon datasets (NDCG@10). Each subplot shows absolute NDCG@10 (bars: SASRec in blue, ReAd in orange) and relative gain (green/red dots with dashed line) across item popularity quartiles (Q1: long-tail, Q4: popular). ReAd achieves substantial improvements on long-tail items (Q1) and moderately popular items (Q2-Q3).}
    \Description{A single row of four charts for Beauty, Home, Office, and Sports compares SASRec and ReAd NDCG at 10 across four popularity quartiles and overlays ReAd's relative gain.}
    \label{fig:item_stratified_ndcg}
\end{figure*}

\begin{figure}[!t]
    \centering
    \includegraphics[width=0.9\linewidth]{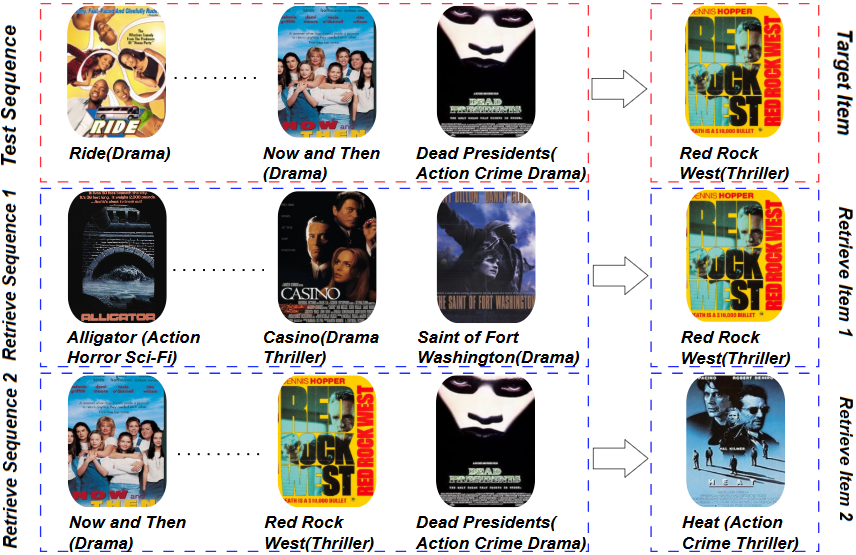}
    \caption{MovieLens case study: red dashed lines denote the test sequence and target item; blue dashed lines denote the two retrieved sequences and corresponding items.}
    \Description{A MovieLens case study shows one test sequence and target linked by red dashed lines and two retrieved historical sequences and items linked by blue dashed lines.}
    \label{fig:case_study}
\end{figure}

\subsection{Item Popularity Stratified Analysis (RQ5)}
\label{sec:pop_analysis}

To answer RQ5 and investigate whether ReAd merely retrieves popular items or genuinely supports long-tail recommendations, we conduct a stratified analysis by item popularity. Following prior work~\cite{long-tail}, we group items into four quartiles (Q1-Q4) by their frequency in the training set, where Q1 contains the least popular (long-tail) items and Q4 the most popular ones. We then evaluate SASRec and ReAd (+SASRec) on each group separately, reporting NDCG@10 as the primary metric.

Figure~\ref{fig:item_stratified_ndcg} presents item popularity stratified results. ReAd consistently improves over SASRec across all popularity quartiles. Notably, substantial gains are observed on long-tail items (Q1), confirming that ReAd does not merely amplify popular items.

The gains follow an inverted-U shape: they are substantial for Q1-Q3 but diminish for the most popular items (Q4). This is because moderately popular items have sufficient yet not excessive collaborative signals in memory, whereas popular items already benefit from abundant training data. These results reveal a fundamental property: test-time retrieval augmentation is most beneficial when collaborative signals are present but not overwhelming. This characterizes the capability boundary of ReAd and, more broadly, of test-time retrieval augmentation in sequential recommendation.

\subsection{Case Study and Efficiency (RQ4\&RQ5)}
The Figure~\ref{fig:case_study} provides a case study on MovieLens, illustrating how ReAd enhances test-time prediction. The test sequence (red) transitions from drama to thriller.

Retrieved sequences (blue) not only share overlapping items such as \textit{Now and Then}, \textit{Dead Presidents}, and \textit{Red Rock West} — confirming collaborative relevance — but also supply complementary items like \textit{Casino} and \textit{Heat}, which reinforce the emerging thriller preference while bridging earlier drama context (e.g., \textit{Alligator}, \textit{Saint of Fort Washington}).

This example highlights two key advantages of ReAd. First, the model does not rely solely on the most recent interactions; instead, it retrieves sequences that are historically similar and collectively reflect both past and emerging interests. Second, retrieved items are not merely popular or globally similar; they are selectively aligned with the local transition in the test sequence (from drama to thriller), enabling the fusion module to refine predictions toward the actual next item, \textit{Red Rock West}. Thus, ReAd enriches test-time representations by dynamically retrieving collaborative signals.

Meanwhile, we also compare inference efficiency on the ML-1M dataset with a batch size. DuoRec is the fastest (\textit{1.06 ms per batch}), followed by TTA (\textit{1.48 ms}). ReAd incurs moderate overhead (\textit{4.76 ms}), but is substantially faster than RaSeRec (\textit{6.20 ms}) while achieving superior accuracy. This confirms that ReAd offers a favorable trade-off between accuracy and efficiency.

%% file: section/conclusion.tex
\section{Conclusion}
\label{sec:con}
In this work, we proposed ReAd, a test-time retrieval augmentation framework for sequential recommendation. To address the challenge of training-test mismatch on sequential patterns, ReAd introduces a collaborative memory to retrieve historically relevant items, a lightweight retrieval learning module that fuses retrieved items into an augmentation embedding, and an entropy-based refinement mechanism that dynamically balances the original prediction with the augmented signal.
Extensive experiments on five benchmark datasets demonstrate that ReAd consistently improves recommendation performance across different backbone architectures. Our fine-grained analysis reveals that ReAd's effectiveness correlates with interaction density and item popularity, achieving consistent gains across all popularity levels, 
with the most pronounced improvements on moderately popular items. Ablation studies validate the contribution of each component, and comparisons of retrieval strategies confirm that similarity-based retrieval is essential. Furthermore, ReAd maintains competitive inference efficiency with moderate overhead. Overall, ReAd provides a practical, model-agnostic solution for enhancing sequential recommendation at test time, with its effectiveness tied to the richness of available collaborative signals.

%% file: section/acknowledgments.tex
\begin{acks}
This work was supported by the Shenzhen Technology University School-level
project (No.20251061020002).
\end{acks}

%% file: section/genai_disclosure.tex
\section*{GenAI Usage Disclosure}
We used GenAI to help polish our writing, such as catching grammar mistakes,
fixing typos, and improving sentence flow. Aside from this light language
editing, we did not use AI for any other part of this paper.

%% file: reference.bib
@inproceedings{ReDA,
  title={A relevant and diverse retrieval-enhanced data augmentation framework for sequential recommendation},
  author={Bian, Shuqing and Zhao, Wayne Xin and Wang, Jinpeng and Wen, Ji-Rong},
  booktitle={Proceedings of the 31st ACM International Conference on Information \& Knowledge Management},
  pages={2923--2932},
  year={2022}
}

@article{Raserec,
  title={Raserec: Retrieval-augmented sequential recommendation},
  author={Zhao, Xinping and Hu, Baotian and Zhong, Yan and Huang, Shouzheng and Zheng, Zihao and Wang, Meng and Wang, Haofen and Zhang, Min},
  journal={arXiv preprint arXiv:2412.18378},
  year={2024}
}

@inproceedings{RA-TTA,
  author       = {Youngjun Lee and
                  Doyoung Kim and
                  Junhyeok Kang and
                  Jihwan Bang and
                  Hwanjun Song and
                  Jae{-}Gil Lee},
  title        = {{RA-TTA:} Retrieval-Augmented Test-Time Adaptation for Vision-Language
                  Models},
  booktitle    = {The Thirteenth International Conference on Learning Representations,
                  {ICLR} 2025, Singapore, April 24-28, 2025},
  publisher    = {OpenReview.net},
  year         = {2025},
}

@inproceedings{TTA,
  title={Data augmentation as free lunch: Exploring the test-time augmentation for sequential recommendation},
  author={Dang, Yizhou and Liu, Yuting and Yang, Enneng and Huang, Minhan and Guo, Guibing and Zhao, Jianzhe and Wang, Xingwei},
  booktitle={Proceedings of the 48th International ACM SIGIR Conference on Research and Development in Information Retrieval},
  pages={1466--1475},
  year={2025}
}

@inproceedings{hstu,
  title={Actions speak louder than words: trillion-parameter sequential transducers for generative recommendations},
  author={Zhai, Jiaqi and Liao, Lucy and Liu, Xing and Wang, Yueming and Li, Rui and Cao, Xuan and Gao, Leon and Gong, Zhaojie and Gu, Fangda and He, Jiayuan and others},
  booktitle={Proceedings of the 41st International Conference on Machine Learning},
  pages={58484--58509},
  year={2024}
}

@inproceedings{GRU4Rec,
author = {Hidasi, Bal\'{a}zs and Karatzoglou, Alexandros},
title = {Recurrent Neural Networks with Top-k Gains for Session-based Recommendations},
year = {2018},
publisher = {Association for Computing Machinery},
address = {New York, NY, USA},
booktitle = {Proceedings of the 27th ACM International Conference on Information and Knowledge Management},
pages = {843–852},
numpages = {10},
location = {Torino, Italy},
series = {CIKM '18}
}

@inproceedings{bert4rec,
author = {Sun, Fei and Liu, Jun and Wu, Jian and Pei, Changhua and Lin, Xiao and Ou, Wenwu and Jiang, Peng},
title = {BERT4Rec: Sequential Recommendation with Bidirectional Encoder Representations from Transformer},
year = {2019},
isbn = {9781450369763},
publisher = {Association for Computing Machinery},
address = {New York, NY, USA},
booktitle = {Proceedings of the 28th ACM International Conference on Information and Knowledge Management},
pages = {1441–1450},
numpages = {10},
location = {Beijing, China},
series = {CIKM '19}
}

@INPROCEEDINGS {sasrec,
author = { Kang, Wang-Cheng and McAuley, Julian },
booktitle = { 2018 IEEE International Conference on Data Mining (ICDM) },
title = {{ Self-Attentive Sequential Recommendation }},
year = {2018},
volume = {},
ISSN = {},
pages = {197-206},
doi = {10.1109/ICDM.2018.00035},
publisher = {IEEE Computer Society},
address = {Los Alamitos, CA, USA},
}

@inproceedings{caser,
author = {Tang, Jiaxi and Wang, Ke},
title = {Personalized Top-N Sequential Recommendation via Convolutional Sequence Embedding},
year = {2018},
isbn = {9781450355810},
publisher = {Association for Computing Machinery},
address = {New York, NY, USA},
booktitle = {Proceedings of the Eleventh ACM International Conference on Web Search and Data Mining},
pages = {565–573},
numpages = {9},
location = {Marina Del Rey, CA, USA},
series = {WSDM '18}
}

@inproceedings{CGN,
author = {Yuan, Fajie and Karatzoglou, Alexandros and Arapakis, Ioannis and Jose, Joemon M. and He, Xiangnan},
title = {A Simple Convolutional Generative Network for Next Item Recommendation},
year = {2019},
isbn = {9781450359405},
publisher = {Association for Computing Machinery},
address = {New York, NY, USA},
booktitle = {Proceedings of the Twelfth ACM International Conference on Web Search and Data Mining},
pages = {582–590},
numpages = {9},
location = {Melbourne VIC, Australia},
series = {WSDM '19}
}

@inproceedings{SR-GNN,
author = {Wu, Shu and Tang, Yuyuan and Zhu, Yanqiao and Wang, Liang and Xie, Xing and Tan, Tieniu},
title = {Session-based recommendation with graph neural networks},
year = {2019},
isbn = {978-1-57735-809-1},
publisher = {AAAI Press},
booktitle = {Proceedings of the Thirty-Third AAAI Conference on Artificial Intelligence and Thirty-First Innovative Applications of Artificial Intelligence Conference and Ninth AAAI Symposium on Educational Advances in Artificial Intelligence},
articleno = {43},
numpages = {8},
location = {Honolulu, Hawaii, USA},
series = {AAAI'19/IAAI'19/EAAI'19}
}

@inproceedings{CadiRec,
author = {Cui, Ziqiang and Wu, Haolun and He, Bowei and Cheng, Ji and Ma, Chen},
title = {Context Matters: Enhancing Sequential Recommendation with Context-aware Diffusion-based Contrastive Learning},
year = {2024},
isbn = {9798400704369},
publisher = {Association for Computing Machinery},
address = {New York, NY, USA},
booktitle = {Proceedings of the 33rd ACM International Conference on Information and Knowledge Management},
pages = {404–414},
numpages = {11},
location = {Boise, ID, USA},
series = {CIKM '24}
}

@inproceedings{meta,
author = {Qin, Xiuyuan and Yuan, Huanhuan and Zhao, Pengpeng and Fang, Junhua and Zhuang, Fuzhen and Liu, Guanfeng and Liu, Yanchi and Sheng, Victor},
title = {Meta-optimized Contrastive Learning for Sequential Recommendation},
year = {2023},
booktitle = {Proceedings of the 46th International ACM SIGIR Conference on Research and Development in Information Retrieval},
pages = {89–98},
}

@article{ttt4rec,
  title={TTT4Rec: A Test-Time Training Approach for Rapid Adaption in Sequential Recommendation},
  author={Yang, Zhaoqi and Wang, Yanan and Ge, Yong},
  journal={arXiv preprint arXiv:2409.19142},
  year={2024}
}

@inproceedings{TTT,
author = {Zhang, Changshuo and Zhang, Xiao and Shi, Teng and Xu, Jun and Wen, Ji-Rong},
title = {Test-Time Alignment with State Space Model for Tracking User Interest Shifts in Sequential Recommendation},
year = {2025},
isbn = {9798400713644},
publisher = {Association for Computing Machinery},
address = {New York, NY, USA},
booktitle = {Proceedings of the Nineteenth ACM Conference on Recommender Systems},
pages = {461–471},
numpages = {11},
location = {
},
series = {RecSys '25}
}

@inproceedings{long-tail,
author = {Liu, Siyi and Zheng, Yujia},
title = {Long-tail Session-based Recommendation},
year = {2020},
isbn = {9781450375832},
publisher = {Association for Computing Machinery},
address = {New York, NY, USA},
booktitle = {Proceedings of the 14th ACM Conference on Recommender Systems},
pages = {509–514},
numpages = {6},
location = {Virtual Event, Brazil},
series = {RecSys '20}
}

@inproceedings{test-time,
  title={Test-time training with self-supervision for generalization under distribution shifts},
  author={Sun, Yu and Wang, Xiaolong and Liu, Zhuang and Miller, John and Efros, Alexei and Hardt, Moritz},
  booktitle={International conference on machine learning},
  pages={9229--9248},
  year={2020},
  organization={PMLR}
}

@article{DTT,
author = {Yang, Xihong and Wang, Yiqi and Chen, Jin and Fan, Wenqi and Zhao, Xiangyu and Zhu, En and Liu, Xinwang and Lian, Defu},
title = {Dual Test-Time Training for Out-of-Distribution Recommender System},
year = {2025},
issue_date = {June 2025},
publisher = {IEEE Educational Activities Department},
address = {USA},
volume = {37},
number = {6},
issn = {1041-4347},
journal = {IEEE Trans. on Knowl. and Data Eng.},
month = mar,
pages = {3312–3326},
numpages = {15}
}

@inproceedings{fuxi-alpha,
author = {Ye, Yufei and Guo, Wei and Chin, Jin Yao and Wang, Hao and Zhu, Hong and Lin, Xi and Ye, Yuyang and Liu, Yong and Tang, Ruiming and Lian, Defu and Chen, Enhong},
title = {FuXi-$\alpha$: Scaling Recommendation Model with Feature Interaction Enhanced Transformer},
year = {2025},
isbn = {9798400713316},
publisher = {Association for Computing Machinery},
address = {New York, NY, USA},
booktitle = {Companion Proceedings of the ACM on Web Conference 2025},
pages = {557–566},
numpages = {10},
location = {Sydney NSW, Australia},
series = {WWW '25}
}

@inproceedings{longer,
  title={Longer: Scaling up long sequence modeling in industrial recommenders},
  author={Chai, Zheng and Ren, Qin and Xiao, Xijun and Yang, Huizhi and Han, Bo and Zhang, Sijun and Chen, Di and Lu, Hui and Zhao, Wenlin and Yu, Lele and others},
  booktitle={Proceedings of the Nineteenth ACM Conference on Recommender Systems},
  pages={247--256},
  year={2025}
}

@article{onerec,
  title={OneRec Technical Report},
  author={Zhou, Guorui and Deng, Jiaxin and Zhang, Jinghao and Cai, Kuo and Ren, Lejian and Luo, Qiang and Wang, Qianqian and Hu, Qigen and Huang, Rui and Wang, Shiyao and others},
  journal={arXiv preprint arXiv:2506.13695},
  year={2025}
}

@inproceedings{empowering,
  title={Empowering Graph Representation Learning with Test-Time Graph Transformation},
  author={Jin, Wei and Zhao, Tong and Ding, Jiayuan and Liu, Yozen and Tang, Jiliang and Shah, Neil},
  booktitle={ICLR},
  year={2023}
}

@inproceedings{rag,
author = {Fan, Wenqi and Ding, Yujuan and Ning, Liangbo and Wang, Shijie and Li, Hengyun and Yin, Dawei and Chua, Tat-Seng and Li, Qing},
title = {A survey on rag meeting LLMs: Towards retrieval-augmented large language models},
booktitle = {Proceedings of the 30th ACM SIGKDD Conference on Knowledge Discovery and Data Mining},
pages = {6491–6501},
year = {2024}
}

@inproceedings{fusing,
  title={Fusing similarity models with markov chains for sparse sequential recommendation},
  author={He, Ruining and McAuley, Julian},
  booktitle={2016 IEEE 16th international conference on data mining (ICDM)},
  pages={191--200},
  year={2016},
  organization={IEEE}
}

@inproceedings{markov,
author = {Rendle, Steffen and Freudenthaler, Christoph and Schmidt-Thieme, Lars},
title = {Factorizing personalized Markov chains for next-basket recommendation},
year = {2010},
isbn = {9781605587998},
publisher = {Association for Computing Machinery},
address = {New York, NY, USA},
booktitle = {Proceedings of the 19th International Conference on World Wide Web},
pages = {811–820},
numpages = {10},
location = {Raleigh, North Carolina, USA},
series = {WWW '10}
}

@inproceedings{mlp,
author = {Zhou, Kun and Yu, Hui and Zhao, Wayne Xin and Wen, Ji-Rong},
title = {Filter-enhanced MLP is All You Need for Sequential Recommendation},
year = {2022},
isbn = {9781450390965},
publisher = {Association for Computing Machinery},
address = {New York, NY, USA},
booktitle = {Proceedings of the ACM Web Conference 2022},
pages = {2388–2399},
numpages = {12},
location = {Virtual Event, Lyon, France},
series = {WWW '22}
}

@article{attention,
  title={Attention is all you need},
  author={Vaswani, Ashish and Shazeer, Noam and Parmar, Niki and Uszkoreit, Jakob and Jones, Llion and Gomez, Aidan N and Kaiser, {\L}ukasz and Polosukhin, Illia},
  journal={Advances in neural information processing systems},
  volume={30},
  year={2017}
}

@inproceedings{s3,
  title={S3-rec: Self-supervised learning for sequential recommendation with mutual information maximization},
  author={Zhou, Kun and Wang, Hui and Zhao, Wayne Xin and Zhu, Yutao and Wang, Sirui and Zhang, Fuzheng and Wang, Zhongyuan and Wen, Ji-Rong},
  booktitle={Proceedings of the 29th ACM international conference on information \& knowledge management},
  pages={1893--1902},
  year={2020}
}

@inproceedings{cl4srec,
  title={Contrastive learning for sequential recommendation},
  author={Xie, Xu and Sun, Fei and Liu, Zhaoyang and Wu, Shiwen and Gao, Jinyang and Zhang, Jiandong and Ding, Bolin and Cui, Bin},
  booktitle={2022 IEEE 38th international conference on data engineering (ICDE)},
  pages={1259--1273},
  year={2022},
  organization={IEEE}
}

@inproceedings{duorec,
  title={Contrastive learning for representation degeneration problem in sequential recommendation},
  author={Qiu, Ruihong and Huang, Zi and Yin, Hongzhi and Wang, Zijian},
  booktitle={Proceedings of the fifteenth ACM international conference on web search and data mining},
  pages={813--823},
  year={2022}
}

@article{SRA-CL,
  title={Semantic Retrieval Augmented Contrastive Learning for Sequential Recommendation},
  author={Cui, Ziqiang and Weng, Yunpeng and Tang, Xing and Zhang, Xiaokun and Liu, Dugang and Li, Shiwei and Liu, Peiyang and He, Bowei and Luo, Weihong and He, Xiuqiang and others},
  journal={arXiv preprint arXiv:2503.04162},
  year={2025}
}

@article{coserec,
  title={Contrastive self-supervised sequential recommendation with robust augmentation},
  author={Liu, Zhiwei and Chen, Yongjun and Li, Jia and Yu, Philip S and McAuley, Julian and Xiong, Caiming},
  journal={arXiv preprint arXiv:2108.06479},
  year={2021}
}

@inproceedings{Diff4rec,
author = {Wu, Zihao and Wang, Xin and Chen, Hong and Li, Kaidong and Han, Yi and Sun, Lifeng and Zhu, Wenwu},
title = {Diff4Rec: Sequential Recommendation with Curriculum-scheduled Diffusion Augmentation},
year = {2023},
isbn = {9798400701085},
publisher = {Association for Computing Machinery},
address = {New York, NY, USA},
booktitle = {Proceedings of the 31st ACM International Conference on Multimedia},
pages = {9329–9335},
numpages = {7},
location = {Ottawa ON, Canada},
series = {MM '23}
}

@inproceedings{equivariant,
  title={Equivariant contrastive learning for sequential recommendation},
  author={Zhou, Peilin and Gao, Jingqi and Xie, Yueqi and Ye, Qichen and Hua, Yining and Kim, Jaeboum and Wang, Shoujin and Kim, Sunghun},
  booktitle={Proceedings of the 17th ACM Conference on Recommender Systems},
  pages={129--140},
  year={2023}
}

@inproceedings{ICL,
  title={Intent contrastive learning for sequential recommendation},
  author={Chen, Yongjun and Liu, Zhiwei and Li, Jia and McAuley, Julian and Xiong, Caiming},
  booktitle={Proceedings of the ACM Web Conference 2022},
  pages={2172--2182},
  year={2022}
}

@inproceedings{mcl,
author = {Qin, Xiuyuan and Yuan, Huanhuan and Zhao, Pengpeng and Fang, Junhua and Zhuang, Fuzhen and Liu, Guanfeng and Liu, Yanchi and Sheng, Victor},
title = {Meta-optimized Contrastive Learning for Sequential Recommendation},
year = {2023},
booktitle = {Proceedings of the 46th International ACM SIGIR Conference on Research and Development in Information Retrieval},
pages = {89–98},
}

@inproceedings{breaking,
  title={Breaking the Bottleneck: User-Specific Optimization and Real-Time Inference Integration for Sequential Recommendation},
  author={Xie, Wenjia and Wang, Hao and Fang, Minghao and Yu, Ruize and Guo, Wei and Liu, Yong and Lian, Defu and Chen, Enhong},
  booktitle={Proceedings of the 31st ACM SIGKDD Conference on Knowledge Discovery and Data Mining V. 2},
  pages={3333--3343},
  year={2025}
}

@inproceedings{ttt++,
author = {Liu, Yuejiang and Kothari, Parth and van Delft, Bastien and Bellot-Gurlet, Baptiste and Mordan, Taylor and Alahi, Alexandre},
title = {TTT++: when does self-supervised test-time training fail or thrive?},
year = {2021},
isbn = {9781713845393},
publisher = {Curran Associates Inc.},
address = {Red Hook, NY, USA},
booktitle = {Proceedings of the 35th International Conference on Neural Information Processing Systems},
articleno = {1669},
numpages = {13},
series = {NIPS '21}
}

@inproceedings{better,
  title={Better aggregation in test-time augmentation},
  author={Shanmugam, Divya and Blalock, Davis and Balakrishnan, Guha and Guttag, John},
  booktitle={Proceedings of the IEEE/CVF international conference on computer vision},
  pages={1214--1223},
  year={2021}
}

@article{rag-survey,
  title={Retrieval-augmented generation for large language models: A survey},
  author={Gao, Yunfan and Xiong, Yun and Gao, Xinyu and Jia, Kangxiang and Pan, Jinliu and Bi, Yuxi and Dai, Yixin and Sun, Jiawei and Wang, Haofen and Wang, Haofen},
  journal={arXiv preprint arXiv:2312.10997},
  volume={2},
  number={1},
  year={2023}
}

@inproceedings{CoRAL,
author = {Wu, Junda and Chang, Cheng-Chun and Yu, Tong and He, Zhankui and Wang, Jianing and Hou, Yupeng and McAuley, Julian},
title = {CoRAL: Collaborative Retrieval-Augmented Large Language Models Improve Long-tail Recommendation},
year = {2024},
isbn = {9798400704901},
publisher = {Association for Computing Machinery},
address = {New York, NY, USA},
booktitle = {Proceedings of the 30th ACM SIGKDD Conference on Knowledge Discovery and Data Mining},
pages = {3391–3401},
numpages = {11},
location = {Barcelona, Spain},
series = {KDD '24}
}

@inproceedings{RALLRec,
author = {Xu, Jian and Luo, Sichun and Chen, Xiangyu and Huang, Haoming and Hou, Hanxu and Song, Linqi},
title = {RALLRec: Improving Retrieval Augmented Large Language Model Recommendation with Representation Learning},
year = {2025},
isbn = {9798400713316},
publisher = {Association for Computing Machinery},
address = {New York, NY, USA},
booktitle = {Companion Proceedings of the ACM on Web Conference 2025},
pages = {1436–1440},
numpages = {5},
location = {Sydney NSW, Australia},
series = {WWW '25}
}

@inproceedings{RAL-CDNet,
  title={Retrieval Augmented Cross-Domain LifeLong Behavior Modeling for Enhancing Click-through Rate Prediction},
  author={Tang, Xing and Yang, Chaohua and Fu, Yuwen and Ao, Dongyang and Li, Shiwei and Lyu, Fuyuan and Liu, Dugang and He, Xiuqiang},
  booktitle={Proceedings of the 31st ACM SIGKDD Conference on Knowledge Discovery and Data Mining V. 2},
  pages={4891--4900},
  year={2025}
}

@article{faiss,
      title={The Faiss library},
      author={Matthijs Douze and Alexandr Guzhva and Chengqi Deng and Jeff Johnson and Gergely Szilvasy and Pierre-Emmanuel Mazaré and Maria Lomeli and Lucas Hosseini and Hervé Jégou},
      year={2024},
      eprint={2401.08281},
      archivePrefix={arXiv},
      primaryClass={cs.LG}
}

@inproceedings{ics,
author = {Qin, Xiuyuan and Yuan, Huanhuan and Zhao, Pengpeng and Liu, Guanfeng and Zhuang, Fuzhen and Sheng, Victor S.},
title = {Intent Contrastive Learning with Cross Subsequences for Sequential Recommendation},
year = {2024},
isbn = {9798400703713},
publisher = {Association for Computing Machinery},
address = {New York, NY, USA},
booktitle = {Proceedings of the 17th ACM International Conference on Web Search and Data Mining},
pages = {548–556},
numpages = {9},
location = {Merida, Mexico},
series = {WSDM '24}
}

@inproceedings{recbole,
  author    = {Lanling Xu and Zhen Tian and Gaowei Zhang and Junjie Zhang and Lei Wang and Bowen Zheng and Yifan Li and Jiakai Tang and Zeyu Zhang and Yupeng Hou and Xingyu Pan and Wayne Xin Zhao and Xu Chen and Ji{-}Rong Wen},
  title     = {Towards a More User-Friendly and Easy-to-Use Benchmark Library for Recommender Systems},
  booktitle = {{SIGIR}},
  pages     = {2837--2847},
  publisher = {{ACM}},
  year      = {2023}
}

@article{adam,
  title={Adam: A method for stochastic optimization},
  author={Kingma, Diederik P},
  journal={arXiv preprint arXiv:1412.6980},
  year={2014}
}

@inproceedings{reviewforbert4rec,
author = {Petrov, Aleksandr and Macdonald, Craig},
title = {A Systematic Review and Replicability Study of BERT4Rec for Sequential Recommendation},
year = {2022},
isbn = {9781450392785},
publisher = {Association for Computing Machinery},
address = {New York, NY, USA},
booktitle = {Proceedings of the 16th ACM Conference on Recommender Systems},
pages = {436–447},
numpages = {12},
location = {Seattle, WA, USA},
series = {RecSys '22}
}
